\documentclass[useAMS,usenatbib]{mn2e}

\usepackage{color}
\usepackage{amssymb}
\usepackage[pdftex]{graphicx}
\usepackage{epstopdf}
\usepackage[draft]{hyperref}
\hypersetup{colorlinks,breaklinks,linktoc=page,linkcolor=black,citecolor=blue   } 
\usepackage[T1]{fontenc} 
\usepackage{ amssymb }
\usepackage{aecompl}
\usepackage{soul}
\usepackage{amsmath}
\usepackage{eso-pic}

\AddToShipoutPictureBG*{%
  \AtPageUpperLeft{%
    \hspace{17cm}%
    \raisebox{-3.5\baselineskip}{%
      \makebox[0pt][l]{\textnormal{DES 2018-0326}}
}}}%

\AddToShipoutPictureBG*{%
  \AtPageUpperLeft{%
    \hspace{15cm}%
    \raisebox{-4.5\baselineskip}{%
      \makebox[0pt][l]{\textnormal{FERMILAB-PUB-19-103-AE}}
}}}%

\title[Stellar mass as cluster mass proxy]{Stellar mass as a galaxy cluster mass proxy: application to the Dark Energy Survey redMaPPer clusters}

\author[Palmese et al.]{
\parbox{\textwidth}{
\Large
A.~Palmese$^{1,2}$\thanks{E-mail: \url{palmese@fnal.gov}}, 
J.~Annis$^{1}$, J.~Burgad$^{3}$, A.~Farahi$^{4}$, M.~Soares-Santos$^{5}$, B.~Welch$^{6}$, M.~da Silva Pereira$^{5}$, H.~Lin$^{1}$, S.~Bhargava$^{7}$, D.~L.~Hollowood$^{8}$, R.~Wilkinson$^{7}$, P.~Giles$^{7}$, T.~Jeltema$^{8,8}$, A.~K.~Romer$^{7}$, A.~E.~Evrard$^{9,10}$, M.~Hilton$^{11,12}$, C.~Vergara Cervantes$^{7}$, A.~Bermeo$^{7}$, J.~Mayers$^{7,7}$, J.~DeRose$^{13,14}$, D.~Gruen$^{13,14,15}$, W.~G.~Hartley$^{2,16}$, O.~Lahav$^{2}$, B.~Leistedt$^{17}$, T.~McClintock$^{18}$, E.~Rozo$^{18}$, E.~S.~Rykoff$^{14,15}$, T.~N.~Varga$^{19,20}$, R.~H.~Wechsler$^{13,14,15}$, Y.~Zhang$^{1}$, S.~Avila$^{21}$, D.~Brooks$^{2}$, E.~Buckley-Geer$^{1}$, D.~L.~Burke$^{14,15}$, A.~Carnero~Rosell$^{22,23}$, M.~Carrasco~Kind$^{24,25}$, J.~Carretero$^{26}$, F.~J.~Castander$^{27,28}$, C.~Collins$^{29}$, L.~N.~da Costa$^{23,30}$, S.~Desai$^{31}$, J.~De~Vicente$^{22}$, H.~T.~Diehl$^{1}$, J.~P.~Dietrich$^{32,33}$, P.~Doel$^{2}$, B.~Flaugher$^{1}$, P.~Fosalba$^{27,28}$, J.~Frieman$^{1,6}$, J.~Garc\'ia-Bellido$^{34}$, D.~W.~Gerdes$^{9,10}$, R.~A.~Gruendl$^{24,25}$, J.~Gschwend$^{23,30}$, G.~Gutierrez$^{1}$, K.~Honscheid$^{35,36}$, D.~J.~James$^{37}$, E.~Krause$^{38}$, K.~Kuehn$^{39}$, N.~Kuropatkin$^{1}$, A.~Liddle$^{40}$, M.~Lima$^{41,23}$, M.~A.~G.~Maia$^{23,30}$, R.~G.~Mann$^{42}$, J.~L.~Marshall$^{43}$, F.~Menanteau$^{24,25}$, R.~Miquel$^{44,26}$, R.~L.~C.~Ogando$^{23,30}$, A.~A.~Plazas$^{45}$, A.~Roodman$^{14,15}$, P.~Rooney$^{7}$, M.~Sahlen$^{46}$, E.~Sanchez$^{22}$, V.~Scarpine$^{1}$, M.~Schubnell$^{10}$, S.~Serrano$^{27,28}$, I.~Sevilla-Noarbe$^{22}$, F.~Sobreira$^{47,23}$, J. Stott$^{48}$, E.~Suchyta$^{49}$, M.~E.~C.~Swanson$^{25}$, G.~Tarle$^{10}$, D.~Thomas$^{21}$, D.~L.~Tucker$^{1}$, P.~T.~P.~Viana$^{50,51}$, V.~Vikram$^{52}$, A.~R.~Walker$^{53}$
\begin{center} (DES Collaboration) \end{center}
}\vspace{0.3cm} \\
{\small\emph{(Affiliations are listed at the end of paper)} }}

\begin{document}
\pubyear{2019}
\maketitle

\label{firstpage}

\begin{abstract}
We introduce a galaxy cluster mass observable, $\mu_\star$, based on the stellar masses of cluster members, and we present results for the Dark Energy Survey (DES) Year 1 observations. Stellar masses are computed using a Bayesian Model Averaging method, and are validated for DES data using simulations and COSMOS data. We show that $\mu_\star$ works as a promising mass proxy by comparing our predictions to X--ray measurements. We measure the X--ray temperature--$\mu_\star$ relation for a total of {$129$} clusters matched between the wide--field DES Year 1 redMaPPer catalogue and \emph{Chandra} and \emph{XMM} archival observations, spanning the redshift range $0.1<z<0.7$. {For a scaling relation which is linear in logarithmic space,} we find a slope of $\alpha = 0.488\pm0.043$ and a scatter in the X--ray temperature at fixed $\mu_\star$ of $\sigma_{{\rm ln} T_X|\mu_\star}= 0.266^{+0.019}_{-0.020}$ for the joint sample. By using the halo mass scaling relations of the X--ray temperature from the Weighing the Giants program, we further derive the $\mu_\star$--conditioned scatter in mass, finding $\sigma_{{\rm ln} M|\mu_\star}= 0.26^{+ 0.15}_{- 0.10}$. These results are competitive with well--established cluster mass proxies used for cosmological analyses, showing that $\mu_\star$ can be used as a reliable and physically motivated mass proxy to derive cosmological constraints.
\end{abstract}


\section{Introduction}\label{sec:intro}
Galaxy  clusters  are fundamental cosmological probes for large galaxy surveys such as the Dark Energy Survey (DES; \citealt{descollaboration}). The estimation of the cosmological parameters from clusters abundance is allowed by the dependence of the dark matter halo mass function on cosmology (\citealt{press}; \citealt{sheth}; \citealt{tinker}), but it requires estimates of cluster total masses from the observables of our galaxy survey. In practice, we seek cluster mass observables (or mass proxies) that tightly correlate with the {total cluster} mass. 
In other words they exhibit a low scatter in total mass at fixed mass proxy (and vice versa).

Several cluster finders are based on the cluster red-sequence (e.g. \citealt{koester}; \citealt{hao}; \citealt{oguri}; \citealt{redpaper}). Amongst those, redMaPPer (\citealt{redpaper}) has been extensively studied and its mass proxy, the richness $\lambda$, calibrated for large photometric surveys such as the Sloan Digital Sky Survey (SDSS) and the DES over the past decade (\citealt{rozo09,lambda,extrinsicscatter,redmappersv,melchior,simet,projection}). On the other hand, there exists broad evidence that the content of clusters includes a non--negligible fraction of bluer, star--forming galaxies that do not follow the red sequence colour--magnitude relation, in particular towards increasing redshift (\citealt{oemler}; \citealt{butcher1}; \citealt{butcher2}; \citealt{Donahue,zhangII}). This effect is known as the Butcher--Oemler effect.
Whether the inclusion of the blue cloud can improve cluster mass estimates for cosmology is a matter of debate (e.g. \citealt{extrinsicscatter}) and depends on the survey characteristics. At higher redshifts, the blue fraction becomes more significant (though {the number count of blue galaxies over all members} can also reach $\sim 30\%$ below redshift $\sim 0.3$; \citealt{zu2}) and the red sequence is not as distinguishable in colour-magnitude space as at lower redshift. In these regimes, the inclusion of the bluer members may play a significant role in cluster abundance studies of DES and other on--going and future photometric surveys (the Large Synoptic Sky Telescope, LSST, \citealt{lsst}; Euclid, \citealt{euclid}) that push towards higher redshifts, $z=1$ and beyond. 

One clear advantage of including blue galaxies in cluster catalogues is in studying cluster properties and their evolution with redshift, in particular the Butcher-Oemler effect and quenching mechanisms. Moreover, cluster finders able to identify also cluster members that do not belong to the red sequence (\citealt{miller05}; \citealt{vt}) already exist. For these reasons, we develop a low--scatter mass proxy for cluster finders that is not red-sequence based. 

Previous works (for example, \citealt{andreon12}) have exploited stellar masses as a possible cluster mass proxy. We here extend those studies by using a larger sample of X--ray clusters for calibration and by complementing the stellar mass estimates with a membership probability scheme presented in a companion paper, \citet{welch}. A feasibility study for stellar mass computation with DES data has already been carried out in \citet{palmese}, where they found that stellar masses of cluster members can be recovered within 25\% of Hubble Space Telescope Cluster Lensing and Supernovae Survey with Hubble (CLASH) values. The use of the stellar mass content as a probe of total cluster mass is empirically but also physically motivated by the stellar--to--halo connection (see e.g. \citealt{2018arXiv180403097W} for a review), which follows a linear relation in the logarithm of masses at cluster scales. An analysis of the scaling relation for {the stellar content} with halo mass thus has interesting implications not only for cosmological analysis, but also for the stellar--to--halo mass relation (SHMR), which is of interest to understand galaxy evolution within clusters (see \citealt{palmese19} for implications on the SHMR from the whole DES redMaPPer sample). In fact, the SHMR can be expressed as a joint likelihood of mass and observable properties. Because of this, the stellar mass can potentially be more tightly related to the total mass of clusters on the individual halo basis, than {galaxy} number counts would. On the other hand, projection effects due to foreground and background galaxies being confused with cluster members, are perhaps one of the most problematic issue in cluster cosmology with richness \citep{projection}. These effects are likely to affect our stellar mass observable in a similar way to $\lambda$, because of the similarities between the two methods. {However, these effects could be alleviated by the fact that the very massive central galaxies tend to dominate the total mass, making the our mass proxy less sensitive to field galaxies contamination (as recently shown in \citealt{Bradshaw}).}

We therefore apply our method to a well--established cluster catalogue, the DES Year 1 (Y1) redMaPPer catalogue, matched to X--ray observations. Nevertheless, this mass proxy can easily be used with other, non--red--sequence based, cluster finders. We also test our cluster stellar mass estimates against simulations. 

The X--ray temperature and luminosity of clusters represent a well--known, low scatter mass proxy for cluster mass, for which total mass scaling relations have been studied in depth (e.g. \citealt{Mantz:2016WtG-V}). The formalism by \citet{E14} allows us to link the scaling relations of different mass proxies when a lognormal covariance is assumed around the mean scaling relations of the mass proxies. It is thus possible to derive an estimate of the scatter on the total mass of clusters by using the scaling relations between our mass proxy and the X--ray temperature, and between the X--ray temperature and the total mass of clusters. Such estimates provide essential prior information for our mass proxy--mass scatter in a cosmological analysis with cluster abundance. 

In this work we present a stellar mass--based cluster mass proxy, called $\mu_\star$, and assess its performance as a mass proxy using archival X--ray data. This paper is divided into five sections. In Section \ref{datasec} we describe the DES galaxy catalogue, the Y1 redMaPPer catalogue and the X--ray cluster catalogues used. In Section \ref{methodsec} we present a new method to compute galaxy stellar masses, that uses a Bayesian model averaging technique. We then introduce the scheme to produce our stellar mass proxy $\mu_\star$ and briefly describe the membership probability assignment. We also present the method used to compute the X--ray scaling relations and the mass scatter. Section \ref{results} contains measurements of the $T_X - \mu_\star$ relation and scatter constraints, both for the temperature scatter and the total cluster mass scatter. Section \ref{sec:conclusion} contains discussion and conclusions.

Throughout this work we assume a $\Lambda$CDM flat cosmology with $H_0=70 ~{\rm km ~s^{-1}~ Mpc^{-1}}$, $\Omega_m = 0.3$, $\Omega_\Lambda =0.7$. The notation adopted for the cluster mass and radius follows the one often used in the literature. The radii of spheres around the cluster centre are written as $r_{\Delta m}$ and $r_{\Delta c}$ where $\Delta$ is the overdensity of the sphere with respect to the mean matter density (subscript $m$) or the critical density (subscript $c$) at the cluster redshift. Masses inside those spheres are therefore $M_{\Delta m}=\Delta \frac{4 \pi }{3}r^3_{\Delta m}\rho_m$ and similarly for $M_{\Delta c}$.  In the following, we quote $\Delta=200$, which roughly corresponds to the density contrast at virialisation for a dark matter halo at $z=0$. Logarithms indicated as ln are in base $e$, and Log are in base 10. Errors are 68\% confidence level unless otherwise stated.

\section{Data}\label{datasec}

\subsection{DES Year 1 data}\label{sec:des}
The DES\footnote{\url{www.darkenergysurvey.org}} is an optical-near-infrared survey that {imaged} 5000 ${\rm deg}^2$ of
the South Galactic Cap in the $grizY$ bands over 575 nights spanning almost six years. The survey {was} carried out using a $\sim3$ $\textrm{deg}^2$ CCD
camera (the DECam, see \citealt{flaugher}) mounted on the Blanco 4-m telescope at the Cerro Tololo Inter-American Observatory (CTIO) in Chile. DES started in 2012 with a testing period (November 2012 -- February 2013) called DES Science Verification (SV)\footnote{For public data release see: \url{http://des.ncsa.illinois.edu/releases/sva1}}. The data used here come from the first year of observations (September 2013 -- February 2014, \citealt{y1}) and cover $1,839~ {\rm deg}^2$ with up to 4 passes per filter. The data are available at \url{http://des.ncsa.illinois.edu/releases/y1a1}.

The survey strategy is designed to optimize the photometric calibration by tiling each region of the survey with several overlapping pointings in each band. This provides uniformity of coverage and control of systematic photometric errors. This strategy allows DES to determine photometric redshifts of $\sim 300$ million galaxies to an accuracy of $\sigma(z) \simeq 0.07 $ out to $z \gtrsim 1$, with some dependence on redshift and galaxy type, and cluster photometric redshifts to $\sigma(z) \sim 0.02$ or better out to $z \simeq 1.3$ (\citealt{descollaboration}). It has already found $\sim 400$ million objects, including stars and galaxies, from the first three years of operations \citep{dr1}, and $\sim 80,000$ galaxy clusters from the first year. 

The DES Data Management (DESDM) pipeline was used for data reduction, as described in detail in \citet{sevilla}, \citet{desai} and \citet{dataproc}. The process includes calibration of the single-epoch images, which are co-added after background subtraction and then cut into tiles. The source catalogue was created using \textsc{Source Extractor (SExtractor}, \citealt{sextractor}) to detect objects on the $riz$ co-added images. 
The median $10\sigma$ limiting magnitudes of Y1 data for galaxies are $g = 23.4$, $r = 23.2$, $i = 22.5$, $z = 21.8$, and $Y = 20.1$.
\citet{firstyear} made further selections to produce a high-quality object catalogue called the Y1A1 ``gold'' catalogue.

\begin{figure}\includegraphics[width=0.5\textwidth]{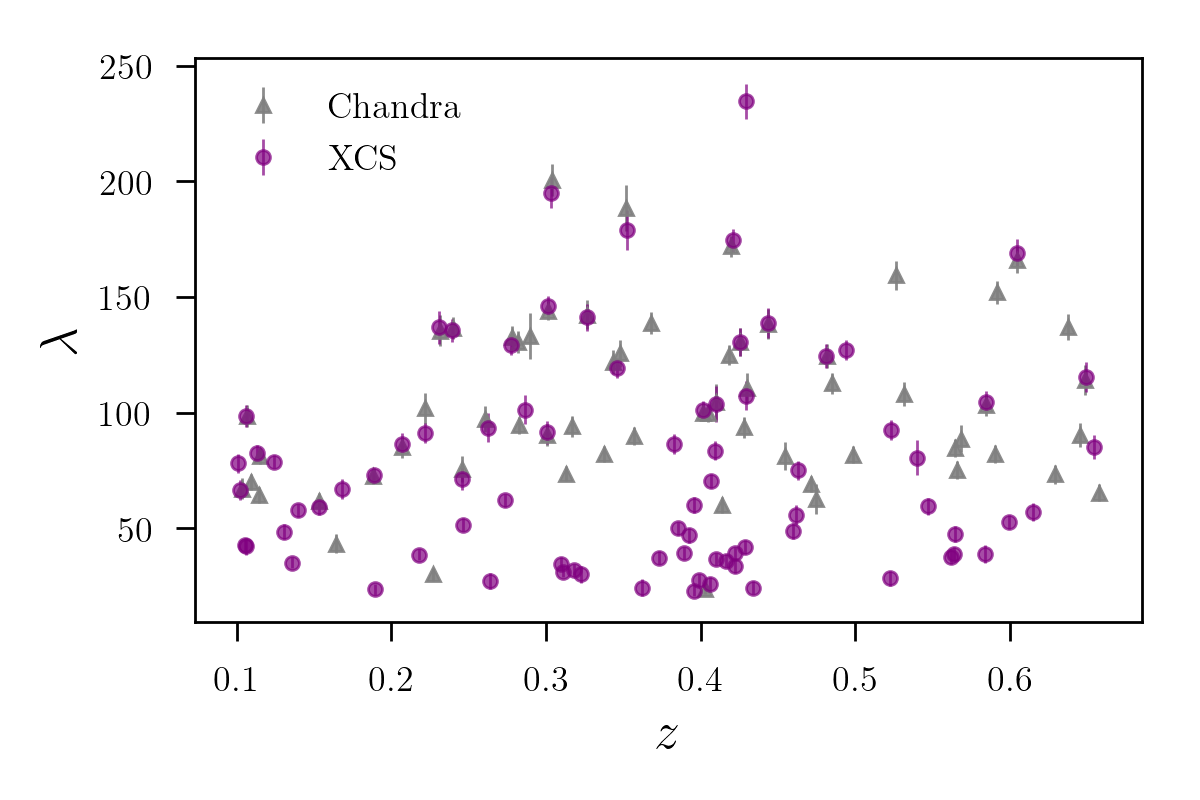}\caption{Distribution in richness $\lambda$ and redshift for the DES Year 1 redMaPPer clusters matched to \emph{Chandra} and \emph{XMM} archival data (\emph{XCS} in the figure) {using the methods presented in \citet{hollowood} and \citet{giles}}.}\label{fig:clusters}\end{figure}

The cluster catalogue used here is the cosmology Y1 redMaPPer catalogue v6.4.14 with richness $\lambda>5$, which consists of 87,297 clusters. 
This sample is then matched to archival X--ray observations from \emph{Chandra} and \emph{XMM}. {We use these X-ray data to measure an X-ray temperature at the position of redMaPPer clusters, rather than cross-matching with existing X-ray cluster catalogs.} The 2D distribution of richness and redshift of the matched samples is shown in Figure \ref{fig:clusters}. The centre position (given by the galaxy with the highest central probability $p_{cen}$) and the cluster redshift are the only outputs used from this catalogue.
The galaxies associated with each cluster are taken from the Y1A1 gold catalogue. We select objects with \texttt{MODEST\_CLASS}=1 in order to exclude sources that are likely not to be galaxies. 

While the cluster catalogue is based on Y1 data, the photometry comes from the deeper Year 3 data (median $10\sigma$ coadded catalogue depths for a $1.95''$ diameter aperture: $g = 24.33$, $r = 24.08$, $i = 23.44$, $z = 22.69$, and $Y = 21.44$ mag; \citealt{dr1}). The photometry  is the result of the Multi-Object Fitting (MOF) pipeline that uses the \texttt{ngmix} code.\footnote{\url{https://github.com/esheldon/ngmix}}

In order to compute the membership probabilities (as described in Section \ref{tomustar}), we use photometric redshifts (photo-$z$'s) from the template-based Bayesian Photometric Redshifts (BPZ) algorithm (\citealt{bpz}). The catalog used in this work uses the same procedure as outlined in \cite{Hoyle}. Briefly, six basic templates taken from \cite{Coleman} and \cite{Kinney} were corrected for redshift evolution and any residual calibration errors. Corrections were performed via finding the best-fit template for a subset of the PRIMUS spectroscopic data set \citep{Cool} and computing median offsets between the observed photometry and template predictions within each template type, in a sliding redshift interval, $\Delta z = 0.06$. The magnitude and galaxy type redshift prior was then calibrated using the COSMOS+UltraVISTA photometric redshift catalogue of \citet{laigle}. Our BPZ run produces redshift probability distributions for $0<z<3.5$ in steps of $dz=0.01$. We use the mean of the probability distribution function (PDF) and an estimate of the width of the PDF: \citet{welch} show that  this is a good enough approximation to estimate membership probabilities with our method. The member galaxy properties are instead computed assuming the much more precise cluster redshift.

\subsubsection{Completeness of the stellar mass sample}

\begin{figure}\includegraphics[width=0.5\textwidth]{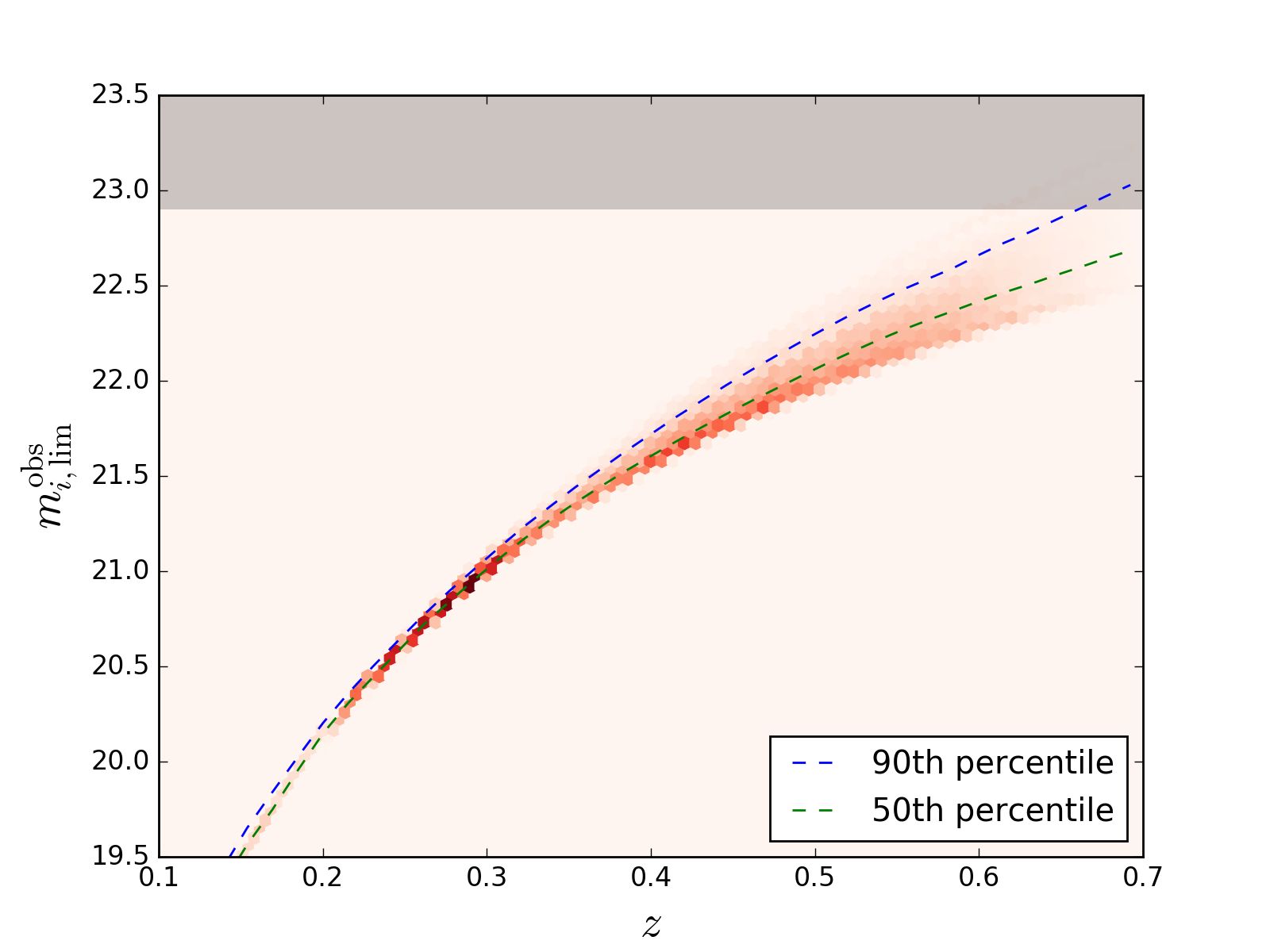}
\includegraphics[width=0.5\textwidth]{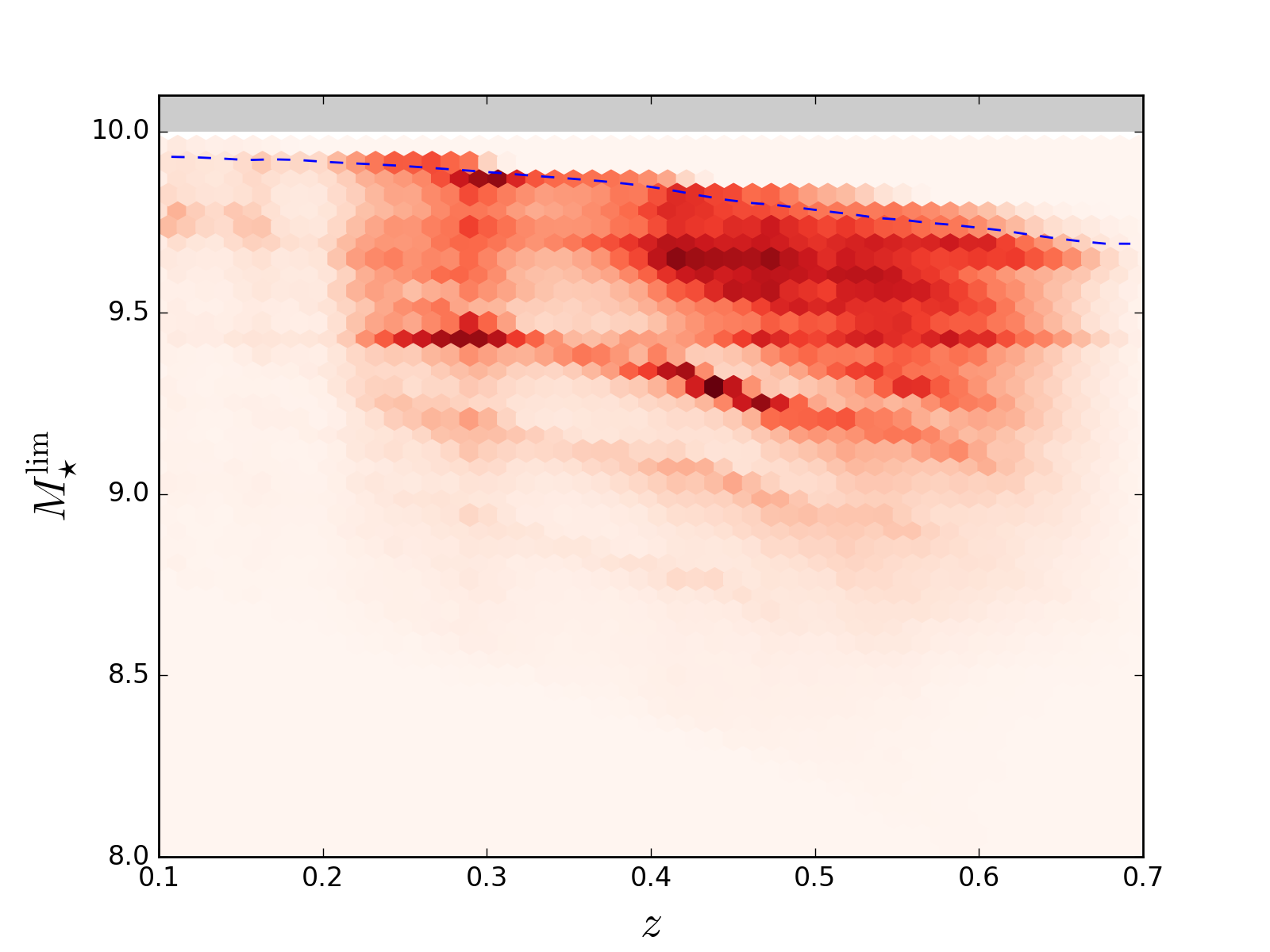}\caption{Analysis of the completeness of the galaxy sample. \emph{Top panel:} observed $i$-band magnitudes that the galaxies in our sample would have if they had the absolute magnitude used as our limit ($M_r^{\rm lim} =-19.8$). The {bottom edge of the} shadowed region represents the DES Y1 95\% completeness limit from \citet{firstyear}.
\emph{Bottom panel:} limiting mass $M_\star^{lim}$ that each galaxy would have, at its redshift, if its absolute magnitude were equal to $M_r^{lim} =-19.8$. The limiting mass is below $10^{10}M_\odot$ at all redshifts; we therefore cut our sample at this stellar mass.
The shadowed region represents this cut. The dashed lines are the 50th and 90th percentile of the distributions. }\label{compl}\end{figure}

The galaxy sample described in Section \ref{sec:des} is cut in $r$-band absolute magnitude. Absolute magnitudes were estimated using K-corrections computed from galaxy templates generated by \texttt{kcorrect} v4.2 (\citealt{blanton}).  We took each galaxy's redshift to be the same as its photo-$z$, found the closest \texttt{kcorrect} template on a grid of redshift and colors ($g-r$, $r-i$, and $i-z$), and used that template's K-correction from observed $i$-band to rest-frame $r$-band to calculate $M_r$.  An absolute magnitude cut $M_r$ brighter than $-19.8$ was then applied to the galaxy catalog. 
This cut ensures that our galaxy sample is volume limited across the redshift range considered. In Figure \ref{compl} we show the observed $r$-band magnitudes that the galaxies in our sample would have if they had an absolute magnitude $M_r =-19.8$ as a function of redshift. These are computed using the K-corrections and distance modulus output by our galaxy Spectral Energy Distribution (SED) fitting code using Bayesian Model Averaging (BMA; described in Section \ref{methodsec}) for galaxies with a membership probability $>15\%$ (corresponding to the median of the membership probability distribution), in order to be representative of a realistic cluster galaxy population. We show that the $90^{{\rm th}}$ percentile of the distribution in redshift bins is below the $95\%$ completeness limit of the DES Y1A1 gold catalog (22.9 in $i-$band) over the redshift range covered by the redMaPPer cosmology catalog. We compare to the Y1 magnitude limit as our galaxy catalog contains objects detected in Y1, even if they are matched to the deeper Y3 photometry. We can conclude that with the chosen cut we are $\sim 90\%$ complete. 

In order to estimate the completeness in stellar mass, we look at the mass $M_\star^{lim}$ each galaxy would have, at its redshift, if its absolute magnitude were equal to $M_r^{lim} =-19.8$. This can be achieved by converting the mass-to-light ratio fitted by BMA through ${\rm Log}(M_\star^{lim}) = {\rm Log}(M_\star) + 0.4(M_r - M_r^{lim})$, where $M_r$ and $M_\star$ are the galaxy estimated absolute magnitude and stellar mass. From Figure \ref{compl} it is clear that, if all the galaxies were at $M_r^{lim}$ or fainter, $\gtrsim 90\%$ of them would have a stellar mass $\lesssim 10^{10} M_\odot$. We therefore are $\gtrsim 90\%$ complete above $M_\star =10^{10} M_\odot$ over the whole redshift range. The scatter in mass at each redshift is given by the scatter in $M/L$ of the different models. We therefore cut our stellar mass sample at $M_\star >10^{10} M_\odot$.

\subsection{X--ray catalogues}\label{sec:xray}

The $\mu_\star$--X--ray mass observable relations are computed using archival \emph{XMM} and \emph{Chandra} data. The DES Y1 redMaPPer cluster catalogue is used to find galaxy clusters in the X--ray databases at the same positions. Consequently, the samples are not X--ray selected. However, X--ray temperature and luminosity measurements are not available for all of the redMaPPer clusters, either due to a lack of archival observation, or the number of  photons are not enough to estimate the luminosity or temperature. {In this work we only focus on temperatures rather than luminosities, since the latter exhibit a larger variance if non-core excised  \citep{Fabian:1994, Mantz:2016WtG-V}. Supplemental survey masks would need to be modeled in order to recover the core-excised measurements. }

The X--ray Multi--Mirror Mission (\emph{XMM}; \citealt{xmm}) is a European Space Agency space mission launched in 1999. The \emph{XMM} Cluster Survey (\emph{XCS}) consists in a search for galaxy clusters in archival \emph{XMM-Newton} observations.  {The DES Y1 redMaPPer sample was matched to all {\em XMM} ObsIDs (with useable EPIC science data) under the requirement that the redMaPPer position be within 13$^{\prime}$ of the aim point of the ObsID.  X-ray sources for each ObsID were then detected using the XCS Automated Pipeline Algorithm \citep[XAPA,][]{LD11}.  At the position of the most likely central galaxy of each redMaPPer cluster, we matched to all XAPA-defined extended sources within a comoving distance of 2~Mpc.  Cutout DES and {\em XMM} images are then produced and visually examined to assign a XAPA source to the optical cluster.}

In order to derive the cluster X--ray temperature, we use the \emph{XCS} Post Processing Pipeline ({\tt XCS3P}) as described in Giles et al. (in prep), and briefly describe the methodology here.  Cluster spectra are extracted and fitted using the {\sc xspec} \citep{Arnaud96} package, performed in the 0.3-7.9 keV band with an absorbed MeKaL model.  The cluster spectra are extracted within $r_{2500c}$, {centered on the XAPA defined center of the cluster emission}, and estimated through an iterative procedure.  An initial temperature is estimated using the XAPA source detection region, and $r_{2500c}$ estimated from the $r_{2500c}$-$kT$ relation of \cite{Arnaud2005}.  {For clusters falling on multiple observations, we utilise all available cameras (i.e. PN, MOS1 and MOS2) in a simultaneous fit, provided the individual camera spectral fits are reliable.  Cameras are only included in the simultaneous fit if the temperature is within the range 0.08 $<$ $T_{X}$ $<$ 30 keV, and contains both upper and lower 1$\sigma$ errors.}  The iteration process is performed until $r_{2500c}$ converged to within 10\%.  {To take into account the background, we used a local background annulus centered on the cluster with an inner and outer radius of 2$r_{2500}$ and 3$r_{2500}$ respectively.  All other detected sources in the field, extended and point sources, were excluded from the analysis.  However, our detection routine does not exclude point sources at the centre of the extended emission.  We do not therefore distinguish these sources from what could be a cool-core, and accept this as part of the intrinsic scatter.  This could present a problem at high redshift where AGN emission could dominate over the emission, but again we accept this as part of the intrinsic scatter.  A recent study in XXL clusters at high redshift (z$\sim$1) determined that this was not significant for robust extended X-ray source detections \citep{Logan19}}.  Furthermore, a calculation of coefficient of variation \citep{Koopmans1964} of $T_{X}$ is performed, defined as the ratio of the standard deviation ($\sigma$) to the mean ($\mu$), given by $C_{v} = \sigma(T_{X})/\mu(T_{X})$.  {The values for $\sigma(T_{X})$ and $\mu(T_{X})$ are taken from the distribution of temperature measurements from all iterations}.  In this work, we adopt a value of $C_{v}<0.25$ as an indicator of a reliable measurement of the iterative temperature, {excluding clusters with $C_{V}>0.25$ from the sample}.  
The final sample is composed of 58 clusters in the DES Y1 wide field{, and the list of clusters is reported in Appendix \ref{app:cluster-cat}.}

The \emph{Chandra X--ray Observatory} is a NASA telescope launched in 1999. In order to obtain X--ray temperatures for archival \emph{Chandra} data, we use the Mass Analysis Tool for \emph{Chandra} ({\tt MATCha}) pipeline, described in \citet{hollowood}. This pipeline finds, downloads, and cleans archival \emph{Chandra} data for each of its input cluster candidates. It then iteratively finds a galaxy cluster center (until converged within 15 kpc), and iteratively fits X--ray temperatures within 500 kiloparsec, $r_{2500c}$, and $r_{500c}$ apertures (until converged within $1\sigma{}$). As with {\tt XCS3P}, {\tt MATCha} performs its fitting using {\sc{xspec}}, with an absorbed MeKaL model. As in  {\tt XCS3P}, {\tt MATCha} performs its fits within the 0.3--7.9 keV band. {For consistency between the \emph{XCS} and \emph{Chandra} samples, we apply to both the same $SNR>5$ cut}. We choose to use temperatures within $r_{2500c}$ for this sample because they are more accurate for nearby clusters, where the the $r_{500c}$ apertures becomes too big compared to the \emph{Chandra} chip. 
Our \emph{Chandra} sample is composed of 64 clusters in the DES Y1 wide field{, and the list of clusters used is reported in Appendix \ref{app:cluster-cat}.}

In order to perform a joint fit between the two X--ray samples and improve our population statistics, we correct for a systematic misalignment between the \emph{Chandra} and \emph{XMM} temperature measurements, as estimated by \citet{redmappersv}:
\begin{equation} \label{eq:temp-scale}
   {\rm Log} ( T_X^{\rm Chandra} ) = 1. 0133 {\rm Log} ( T_X^{\rm XMM} ) + 0.1008\,,
\end{equation}
where the temperatures are in keV. In the following, we use eq. (\ref{eq:temp-scale}) to convert \emph{XMM} temperatures.
{The calibration relation found in \citet{farahi} using the same cluster sample used here (apart from a few clusters falling in the Supernovae fields, which we did not include in this analysis), has 15 clusters with both \emph{Chandra} and \emph{XMM} temperatures, and is consistent with the relation reported above. In our joint analysis, we use the Chandra temperatures, and convert the XMM ones to the Chandra frame for the remaining clusters using Eq. (\ref{eq:temp-scale}).}

\section{Method}\label{methodsec}

\subsection{Stellar mass estimation}

\subsubsection{Stellar mass with Bayesian Model Averaging}
A major cause of uncertainty in stellar mass estimation from broadband photometry is in the model assumptions (see e.g. \citealt{mitchell}) that  are needed in model fitting techniques. These assumptions mainly involve redshift, star formation history (SFH), the initial mass function (IMF), the dust content and the knowledge of stellar evolution at all stages. 
\citet{redmappersv} showed that  the redMaPPer photometric redshifts for DES  are excellent, with errors of the order $\sigma_z/(1+z)\sim 0.01$ up to $z\sim 0.9$. This allows us to safely assume the cluster redshift for the  cluster members and to avoid exploring the photo--$z$ dependence of stellar masses, as was done in another DES study by \citet{capozzi}.
Despite the fact that in the present work we can safely  assume that the  cluster redshift is a good estimate of the real galaxy redshift, all the other  assumptions remain unconstrained. We therefore choose not to ignore the uncertainty on model selection and use a set of robust, up-to-date stellar population synthesis (SPS) models and average over all of them, marginalizing over the model uncertainty. The method used here is called Bayesian Model Averaging (BMA, see e.g. \citealt{hoeting}). BMA has already been successfully applied to galaxy SED fitting parameter estimation in \citet{taylor}.

Our code can be used to estimate physical parameters of galaxies (stellar masses, specific star formation rates, ages, metallicities) as well as cluster stellar masses and total star formation rate (SFR) when provided with cluster membership probabilities, and it is publicly available at \url{https://github.com/apalmese/BMAStellarMasses}.

The BMA  starting point is Bayes' theorem, through which we can write the posterior probability distribution $p(\bar{\theta}_k|D,M)$ of the set of parameters $\bar{\theta}_k$ given the data $D$ and the model $M_k$:
\begin{equation}
p(\bar{\theta}_k|D,M_k)=\frac{p(D|M_k,\bar{\theta}_k)p(\bar{\theta}_k|M_k)}{p(D|M_k)}\,,
\end{equation}
where  $p(D|M_k,\bar{\theta}_k)$ is the likelihood, $p(\bar{\theta}_k|M_k)$ is the prior probability of the parameters given the model $M_k$, and  $p(D|M_k)$ is the evidence. In our case, the set of parameters $\bar{\theta}_k$ define the stellar population properties (e.g. stellar mass, SFH parameters, metallicity) of model $M_k$, and the data $D$ are the galaxy's observed magnitudes.

The model averaged posterior distribution of the parameters $\bar{\theta}_k$ is given by the sum of the single model $M_k$ posteriors, weighted by the model prior:
\begin{equation}
p(\bar{\theta}_k|D)=\frac{\sum_kp(\bar{\theta}_k|D,M_k) p(M_k)}{\sum_k p(M_k|D)}\,.
\end{equation}
From BMA it also follows that the  posterior distribution of a  quantity $\Delta$ is the average of the single model posteriors  for that quantity,  weighted by their posterior model probability:
\begin{equation}
p(\Delta|D)=\sum_kp(\Delta|D,M_k) p(M_k|D)\,.\label{bmaeq}
\end{equation}
The posterior model probabilities can be computed by:
\begin{equation}
p(M_k|D)=\frac{p(D|M_k) p(M_k)}{\sum_k p(D|M_k) p(M_k)}\,,
\end{equation}
where
\begin{equation}
p(D|M_k)=\int p(D|M_k,\bar{\theta}_k)p(\bar{\theta}_k|M_k){\rm d}\bar{\theta}_k\,.\label{bmaeq}
\end{equation}
In our case $p(\bar{\theta}_k|M_k)$ is simply a delta function, as the parameters  $\bar{\theta}_k$ (i.e. the SFH parameters, metallicities, etc.) fully  define our models $M_k$.

From Eq. (\ref{bmaeq}) one can write:
\begin{equation}
\langle\Delta\rangle=\sum_k \bar{\Delta}_k p(M_k)\mathcal{L}_k\, , \label{meaneq}
\end{equation}
where $\bar{\Delta}_k$ is the mean $\Delta$ value from the model $M_k$, which is defined by the set of parameters $\bar{\theta}_k$ including metallicity and SFH parameters. $\mathcal{L}_k$ is the likelihood $p(D|M_k)$ that we will reconstruct from the $\chi^2$ distribution. {The model prior $p(M_k)$ is uniform over all models.}

In our code, the mass-to-light ratio $M_\star/L$ is the quantity $\Delta$. Its posterior mean over all the models considered is then used to estimate the stellar mass $M_\star$ of each single galaxy through:
\begin{equation}
{\rm Log}{(M_\star/M_{\odot})}=\langle M_\star/L \rangle-0.4(i-DM+\langle kii \rangle -4.58)\,,
\end{equation}
where $\langle M_\star/L \rangle$ is the weighted mean stellar-mass-to-light-ratio in solar mass units, $i$ is the observed $i$ band magnitude, $DM$ is the distance modulus,  $\langle kii \rangle$ is the weighted mean of the K-correction $i_{\rm rest frame}-i$ and 4.58 is the $i$-band absolute magnitude of the Sun. Weighted means are considered over all models.

In this work we use the flexible stellar population synthesis (FSPS) code by \citet{fsps} to generate  simple  stellar population spectra. Those are computed assuming Padova (\citealt{padova1}, \citealt{padova2}, \citealt{padova3}) isochrones  and Miles (\citealt{miles}) stellar libraries with four different metallicities ($Z=0.03,0.019,0.0096$ and 0.0031). We choose the four-parameter SFH described in \citet{simha}:
\begin{equation}
SFR(t)= \begin{cases}
 A(t-t_i){\rm e}^{(t-t_i)/\tau} & {\rm if }\; t<t_t\\
 SFR(t_t)+\Gamma(t-t_t) & {\rm otherwise}
\end{cases}
\end{equation}
where $t_i$ is the time at which star formation commences ($\sim 1$ Gyr), $t_t$ is the time when the SFR transitions from exponential to a linear fall off ($\sim 9$ Gyr), $\tau$ is the exponential time scale, and $\Gamma$ is the slope of the linearly decreasing SFR as a function of time $t$ after $t_t$. Defining $\theta$ as $\Gamma\equiv{\rm tan}\theta$, we make the  four parameters vary on a grid of values within the following ranges: $\tau\in [0.3,13]$ Gyr, $t_i \in [0.7,2]$ Gyr, $t_t \in [7,13]$ Gyr, and $\theta\in [-10,-80 ]$ deg. Table \ref{tab:sed} reports the grid of values used for these parameters.

\begin{table}\centering
\begin{tabular}{cc}
\hline
\hline
Parameter & Values \\
\hline
$Z$& 0.03, 0.019, 0.0096, 0.0031\\
$t_i$ [Gyr]& 0.7, 1.0, 1.5, 2.0  \\
$t_t$ [Gyr]& 7, 9, 11, 13 \\
$\tau$ [Gyr]& 0.3, 1.0, 1.3, 2.0, 9.0, 13.0 \\
$\theta$ [deg]& $-10$, $-20$, $-30$, $-40$, $-50$, $-80$ \\
\end{tabular}\caption{Parameters of the models used in the BMA SED fitting.}\label{tab:sed}
\end{table}

For each observed galaxy we construct the likelihood $\mathcal{L}_k$ in Eq. (\ref{meaneq}) as $\mathcal{L}_k\propto e^{-\chi^2_k/2}$, with:
\begin{equation}
    \chi^2_k=\sum_j\frac{(C_i-C_{k,j})^2}{\sigma_{C_j}^2} \, ,
\end{equation}
 where the sum is over the colours $g-r$, $r-i$, and $i-z$. $C_j$ are the observed colours, while $C_{k,j}$ are the colours predicted by the model $M_k$ for the colour $j$. The scaling for the theoretical model is given by the $i$ band filter. $\sigma_{C_j}$ are the observed errors added in quadrature with a lower limit of 0.02. 

\subsubsection{Validation of the BMA method}

\begin{figure}\includegraphics[width=0.5\textwidth]{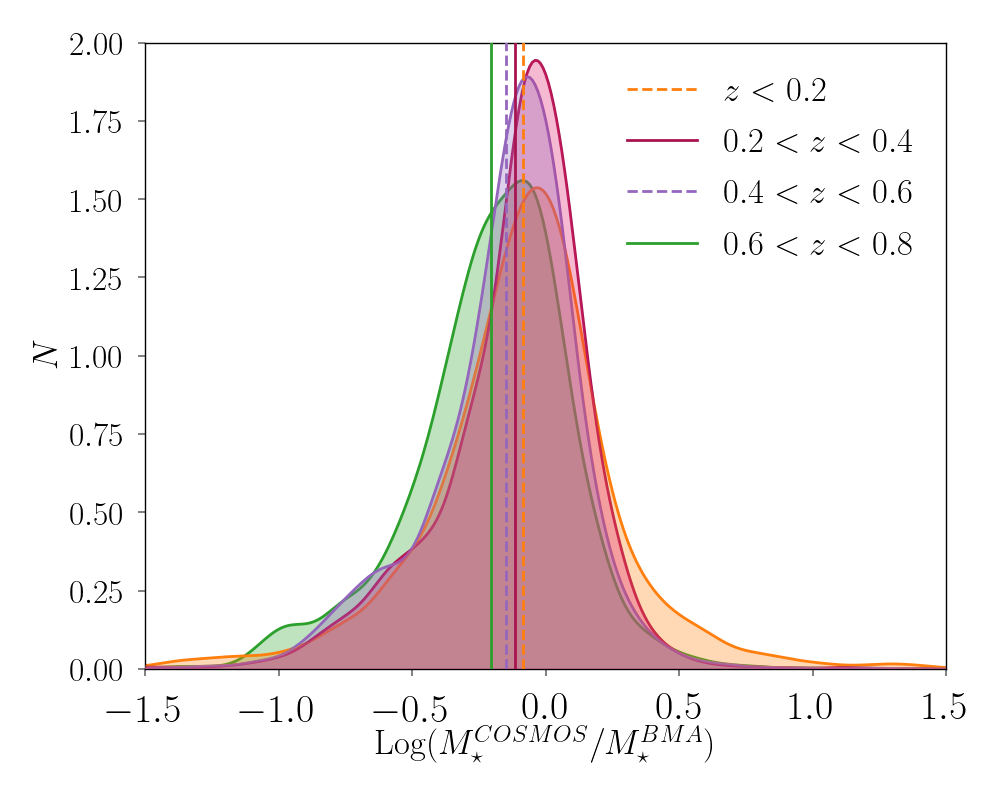}\caption{Comparison of galaxy stellar masses from \citet{laigle} using COSMOS data with those computed with the BMA algorithm using DES data in different redshift bins. The lines represent the mean value of the distributions with the same colour. The histograms have been smoothed with a Gaussian kernel for visualization purposes, and arbitrarily renormalised. The total number of galaxies used is $\sim 120,000$. }\label{fig:cosmos}\end{figure}

\begin{figure}\includegraphics[width=0.5\textwidth]{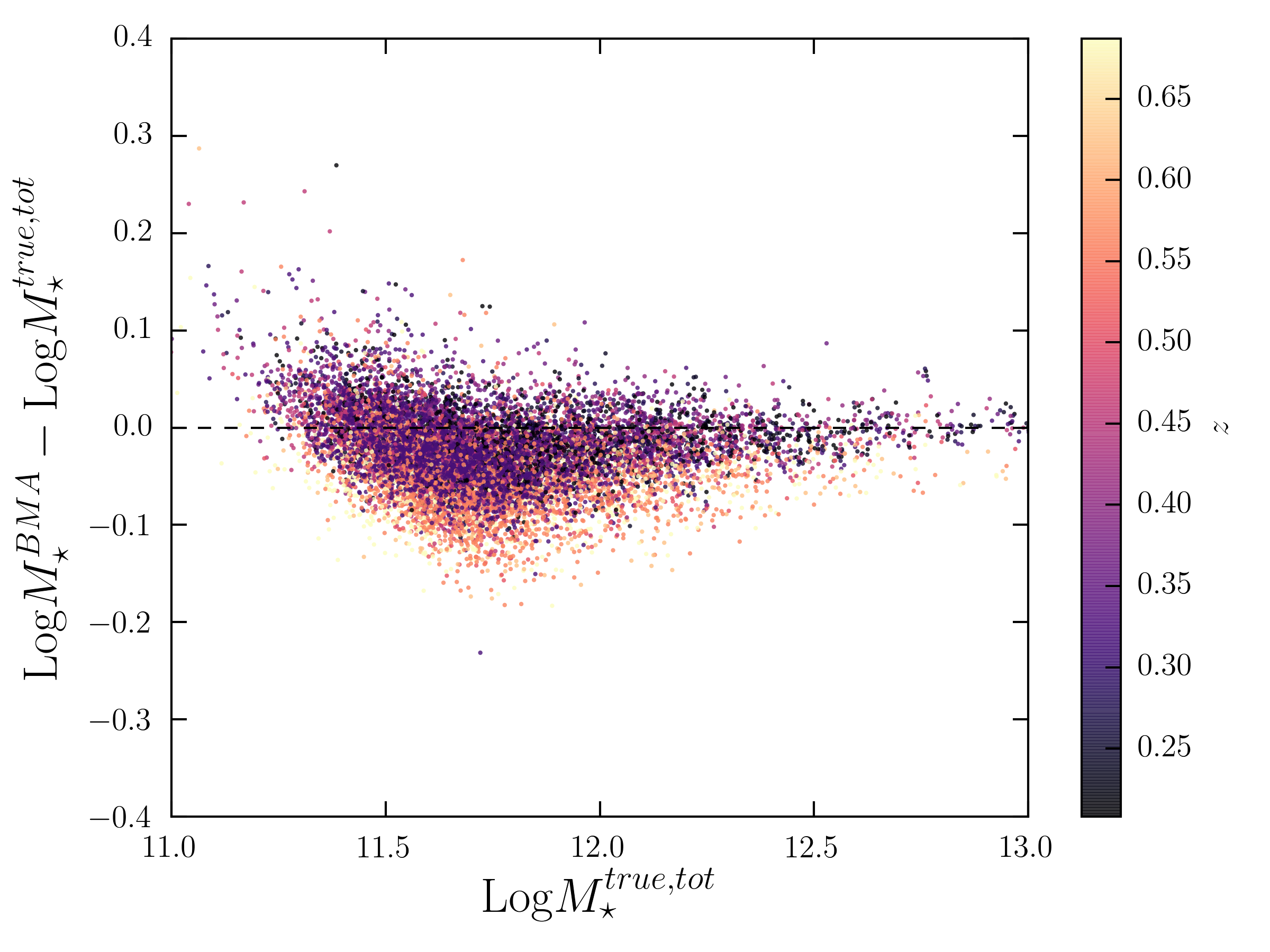}\caption{Comparison of BMA clusters stellar mass to Millennium simulation true values at different redshifts. The dashed lines indicates no difference between the BMA estimates and the true values.}\label{fig:sims}\end{figure}

In order to test our method for stellar mass estimation, we use as reference a catalog that overlaps with DES observations. \citet{laigle} used \textsc{LePhare} to compute stellar masses with multiband data in 16 filters from UV to infrared over the $2~{\rm deg}^2$ COSMOS field. From this sample, matched to DES data, we remove all objects at $z=0$ to eliminate stars, and at $z>1.5$, as higher redshift galaxies are beyond the interests of this work.  Galaxies with  $i$-band  magnitude above 23.0 are also cut out. The remaining sample comprises galaxies with $SNR>10$ in DES, for which we compute stellar masses using the BMA code and DES data. The bias distribution given by the difference in log galaxy stellar mass between the two samples ${\rm Log}(M_\star^{\rm COSMOS})-{\rm Log}(M_\star^{\rm BMA})$ is shown in Figure \ref{fig:cosmos}. Mean bias and scatter (that we quantify as the standard deviation of the distribution) are below the typical error on galaxy stellar masses from SED fitting ($\sim 0.2$ dex) in the redshift range $0.2<z<0.6$, where we expect good performance for optical surveys such as DES. At higher redshift, it is harder to constrain the optical to near-infrared (NIR) SED with the DES bands and therefore the scatter increases. Also at low redshifts ($z<0.2$), the 4000~\AA~break is harder to constrain, as it is blue-ward of the $g$-band effective wavelength. {The $\sim 0.1$ dex bias that can be seen from Figure 3, particularly towards higher redshift, is mostly due to the degeneracies between stellar mass and dust extinction. In fact, we find that the bias is almost null for passive, mostly dust-free galaxies, while it is more pronounced for star--forming dusty galaxies. 
Because our set-up does not include dust, the resultant masses are biased high (because the presence of dust makes them redder). Including dust in our models would introduce further systematics since our wavelength coverage does not extend to restframe infrared wavelengths, particularly at higher redshifts.} \citet{laigle} are able to constrain dust extinction because of the information brought by the infrared data available to them. {Overall, differences between the two catalogues will also be partially due to the fact that the COSMOS stellar masses are not ``true'' stellar masses, and will depend on the assumptions and methodology in \citet{laigle}. Among those assumptions, the synthetic templates assumed by \citet{laigle} are from \citet{bc03}, and thus will differ from the models assumed here.} {We conclude that the observed bias is due to a choice of  templates, rather than to the BMA method itself.}

{One of the main advantages of the BMA method is that it allows to formally incorporate the model uncertainty into the stellar mass uncertainty. This is particularly important for star-forming galaxies. As it turns out from the discussion above, stellar mass estimates of red and passive galaxies are found to be more robust. We find that the uncertainties derived from BMA for this type of galaxy are comparable to those from a more standard approach such as the one adopted in \textsc{LePhare}, using a similar set of templates and the same input magnitudes. On the other hand, blue star-forming galaxies uncertainties are larger by a factor 1.5-2, reflecting the fact that there are a number of models that could similarly fit that data.}

We also test our results against the Millennium simulation semi--analytic model from \citet{delucia}\footnote{\url{http://gavo.mpa-garching.mpg.de/Millennium/Help?page=databases/millimil/delucia2006a}}, and show the results for the sum of stellar mass in clusters {(selected as halos with $M_{200c}>10^{14}~{ M_{\odot}}$)}, in Figure \ref{fig:sims}. We run the BMA algorithm using the simulated magnitudes for the $griz$ SDSS filters, which are very similar to the DES ones. In this case the scatter of the bias distribution is even lower (standard deviation is $\sim 0.04$ dex) than what found in the comparison with the COSMOS results, showing that our method works well against other SED fitting methods and simulations. 



\subsection{From galaxy stellar masses to $\mu_\star$}\label{tomustar}

The cluster mass proxy $\mu_\star$ is computed by weighing the stellar mass of each galaxy in the cluster by its membership probability $p_{{\rm mem}, i}$:
\begin{equation}
\mu_\star = 10^{-10} M^{-1}_\odot~\sum_i  p_{{\rm m}, i} M_{\star,i}\,,
\end{equation}
where the factor $10^{-10}$ simply gives to the mass proxy an order of magnitude similar to that of the number of observed cluster galaxies. The sum is over all the galaxies from the DES Y1A1 gold catalog having $M_r<-19.8$ and within 3 Mpc from the centre of the cluster as given by the redMaPPer Y1 catalog. {The membership probability is computed as described in \citet{welch}, where the membership assignment scheme is presented in detail, together with measurements of the red sequence for redMaPPer clusters.} The probability is given by
\begin{equation}
p_{\rm m} = p_{R}\,p_{z}\,,\label{eq:pmem}
\end{equation}
where the components represent the probability of the galaxy being a member given its redshift ($p_z$) and its distance from the cluster center ($p_R$). The radial probability $p_R$ is assigned by assuming a projected Navarro--Frenk--White (NFW; \citealt{nfw}) profile, with $R_{200c}$ computed by counting galaxies within 3 Mpc and finding the halo profile by assuming an {Halo Occupation Distribution} (HOD) model. {The redshift probability $p_z$ is computed by comparing the photo--$z$ $p(z)$ of single galaxies to the cluster redshift. The membership probability presented here differs from the one provided by redMaPPer because it uses photometric redshift information, instead of a red sequence calibration. The computation of the radial probability is similar, as it assumes the same function for the galaxy profile, while the radius may be different as our method utilises the HOD model.}

We also provide a colour probability $p_{\rm c}$, {the probability that a galaxy belongs to either the red sequence or the blue cloud given its colour. This is} estimated through a Gaussian Mixture Model (GMM) similar to \citet{Hao2009PRECISIONMODEL}. This method fits two Gaussians to the colour distribution of the galaxies in each cluster, weighted by their radial and redshift probabilities. The Gaussians fit the colour distribution of the red sequence and blue cloud of cluster galaxies well. The area of the Gaussians $w_{\rm red}$ and $w_{\rm blue}$ satisfies $w_{\rm red}+w_{\rm blue}=1$ and is used to compute the colour probability:
\begin{equation}
p_{\rm c} = w_{\rm red}p_{\rm red}+w_{\rm blue}p_{\rm blue}\, ,\label{eq:pc}
\end{equation}
where $p_{\rm red}$ ($p_{\rm blue}$) is the probability that a galaxy belongs to the red  sequence (blue cloud) given its colour and the GMM estimates.
The cluster colour distribution is derived after a local background subtraction is performed by measuring the colour distribution of galaxies in the cluster outskirts (between 3 and 5 Mpc). This is done for the colours $g-r$, $r-i$, and $i-z$. The inclusion of the colour probability {as an extra term multiplied to the right-hand-side of Eq. (12)} is tested in Section \ref{results}.
See \cite{welch} for a full description of the membership probability scheme.

The errors on $\mu_\star$ were computed using jackknife resampling. Intuitively, this method allows us to estimate the variance on our estimator by considering a galaxy cut from the cluster at each time.

{We note that the impact of the stellar mass bias seen in Figure 3 for COSMOS galaxies does not have a significant effect on $\mu_\star$. In fact, the bias is reduced to 0.05 dex for a sample of galaxies such as those used in this work for Y1 clusters ($M_\star>10^{10}~M_\odot$, $z<0.7$), with most of the contribution coming from star-forming galaxies. If we assume that the star-forming galaxies can constitute up to $\sim 30\%$ of cluster galaxies in number, and that their typical masses are $10^{10.5}~M_\odot$ (versus $10^{10.7}~M_\odot$ of the passive galaxies, as computed from the COSMOS $M_\star>10^{10}~M_\odot$, $z<0.7$ sample), the maximum bias introduced on $\mu_\star$ is 2.6\%, which is well below the typical 10\% errors in $\mu_\star$.\footnote{Note that up to $\sim 30\%$ of cluster galaxies are blue \citep{zu2}, but not necessarily star-forming. Even if most blue galaxies are star-forming compared to red sequence galaxies, there exist a smaller fraction of blue galaxies which are passive, and red galaxies that are star-forming (e.g.\citealt{2009MNRAS.400..687M}). However, even if we assumed a conservative fraction of 50\% for the content of star-forming galaxies in clusters, the expected upper limit ($\sim4.7\%$) of the bias would still be below our uncertainties.} In reality, this upper bound is further reduced by the radial membership probabilities $p_R$, because the passive, red sequence galaxies dominate in the center of clusters (e.g. \citealt{2009MNRAS.400..687M}). }

\subsection{Hot gas temperature -- stellar mass relations}

Following previous works (e.g. \citealt{rozo09}; \citealt{extrinsicscatter}; \citealt{E14}; \citealt{mulroy2019}; \citealt{farahi}), we assume that the likelihood of a cluster to have {a true value of the X--ray temperature $T^{\rm tr}$, given that it has a stellar mass true value $\mu_\star^{\rm tr}$}, is a log-normal function. Following the notation in \citet{E14} and \citet{farahi}:
\begin{equation}
    P(T^{\rm tr}|\mu_\star^{\rm tr},z)=\frac{1}{\sqrt{2\pi}\sigma_{{\rm ln} T|\mu_\star}} \exp{\Big[- \frac{({\rm ln}~T^{\rm tr}-\langle {\rm ln}~ T^{\rm tr}|\mu_\star^{\rm tr},z\rangle)^2}{2\sigma^2_{{\rm ln} T|\mu_\star}}} \Big]\, ,
\end{equation}
where $\sigma_{{\rm ln} T|\mu_\star}$ is the intrinsic scatter of the hot gas temperature at fixed stellar mass, $\mu_\star^{\rm tr}$. We perform a Bayesian linear regression \citep{kelly} to estimate a linear relation between the logarithm of the X--ray temperature and the logarithm of stellar mass. The free parameters include normalization, slope, and the scatter about the mean relation. Namely, we fit:
\begin{equation}
\langle {\rm ln}~ T|\mu_\star,z\rangle = \big[ \pi_{T|\mu_\star} +\frac{2}{3}{\rm ln}(E(z))\big]+\alpha_{T|\mu_\star}~{\rm ln}\Big( \frac{\mu_\star}{\tilde{\mu}_\star} \Big)\, ,\label{eq:tx}
\end{equation}
where  $\tilde{\mu}_\star$ is the median $\mu_\star$ of the sample and $E(z)=H(z)/H_0$ is the Hubble parameter evolution. The normalization term containing $E(z)$ takes into account the redshift dependence of the temperature, as expected for a self--similar evolution of the intra-cluster medium in virial equilibrium (\citealt{Kaiser:1991,BryanNorman:1998}). We use the publicly available Python version of \citet{kelly} to perform this regression, which provides samples from the posterior distribution of the normalization, $\pi_{T|\mu_\star}$, slope, $\alpha_{T|\mu_\star}$, and scatter about the mean relation, $\sigma_{{\rm ln} T|\mu_\star}$. {The regression code takes into account uncertainties associated with both the dependent and independent variables by assuming a mixture model. Such errors are assumed to be lognormal, and they are transformed to lognormal space through $\sigma_{\ln T}=\sigma_{T}/T$.}

\subsection{Mass scatter inference}

We follow \citet{E14} and \citet{farahi} model to infer the halo mass scatter at fixed $\mu_\star$. \citet{E14} proposed a population model which computes a closed form solution for conditional properties of an observable, here $T_X$, given a selection observable, here stellar mass $\mu_\star$, as a function of their halo properties. We employ their model to infer the halo mass scatter.
In the following, we denote the log of halo mass by $\ln M$. According to their population model, the scatter in temperature at fixed $\mu_\star$ can be written as:
\begin{equation}
\frac{\sigma^2_{{\rm ln}T|\mu_\star}}{\alpha^2_{T| M}} = 
\left[ \sigma^2_{\ln M|\mu_\star}+\sigma^2_{\ln M| T } - 2r_{\mu_\star T}\sigma_{\ln M|\mu_\star}\sigma_{\ln M|T}  \right] .
\label{eq:varT}
\end{equation} 
We employ $\sigma_{\ln M|T} = \sigma_{T|\ln M} / \alpha^2_{T| M}$ and solve for $\sigma_{\ln M | \mu_\star}$, which is the quantity of interest. After rearrangement, the solution yields
\begin{equation}
\sigma^2_{\ln M|\mu_\star} =  \sigma^2_{\ln M|T} \, 
\left[ \left(\frac{\sigma^2_{{\rm ln T}|\mu_\star}}{\sigma^2_{\ln T | M}} - (1-r^2_{\mu_\star T}) \right)^{1/2} + r^2_{\mu_\star T}  \right]^2  ,
\label{eq:varmulambda}
\end{equation} 
where $r_{\mu_\star T}$ is the correlation coefficient between $\mu_\star$ and temperature deviations about their mean values at fixed halo mass. {The values used for $r_{\mu_\star T}$ and $\sigma^2_{\ln T | M}$ will be discussed in Section 4.3.}

\section{Results and discussion}\label{results}
\begin{table*}\centering
\begin{tabular}{c|cccc}
\hline
\hline
 Sample &$\pi_{{\rm ln} T_X|\mu_\star}$&$\alpha_{{\rm ln} T_X|\mu_\star}$&$\sigma_{{\rm ln} T_X|\mu_\star}$ & ${\rm ln}(\tilde{\mu}_\star)$\\
 \hline
\emph{XMM} & $1.306^{+0.035}_{-0.035}$ & $0.449^{+0.054}_{-0.055}$ & $0.277^{+0.026}_{-0.029}$ & 6.61\\
\emph{Chandra} & $1.887^{+0.032}_{-0.032}$ & $0.463^{+0.072}_{-0.072}$ & $0.229^{+0.025}_{-0.027}$ & 7.07\\
\emph{Chandra+XMM} & $1.711^{+0.024}_{-0.025}$ & $0.488^{+0.043}_{-0.043}$ & $0.266^{+0.019}_{-0.020}$& 6.85\\
\hline
\emph{Chandra+XMM} ~$(z<0.3)$ & $1.781^{+0.041}_{-0.042}$ & $0.501^{+0.084}_{-0.084}$ & $0.262^{+0.029}_{-0.033}$& 6.79\\
\emph{Chandra+XMM} ~$(0.3<z<0.5)$ & $1.753^{+0.035}_{-0.036}$ & $0.497^{+0.056}_{-0.055}$ & $0.263^{+0.027}_{-0.031}$& 7.03\\
\emph{Chandra+XMM} ~$(z>0.5)$ & $1.682^{+0.066}_{-0.065}$ & $0.54^{+0.13}_{-0.14}$ & $0.311^{+0.051}_{-0.064}$& 6.91\\
\end{tabular}\caption{Scaling relation parameters from this work following Eq.(\ref{eq:tx}) for X--ray temperatures. {The upper part of the table shows our results for the temperature over the whole $0.1<z<0.7$ redshift range. The bottom section presents results in different redshift bins for the temperatures from the joint \emph{Chandra+XMM} sample.} 
Values represent the median of the parameters posterior distribution, and the errors are the 16th and 84th percentiles. Temperatures are in units of keV.
}\label{tab:tx}
\end{table*}

\begin{figure*}
\includegraphics[width=0.45\textwidth]{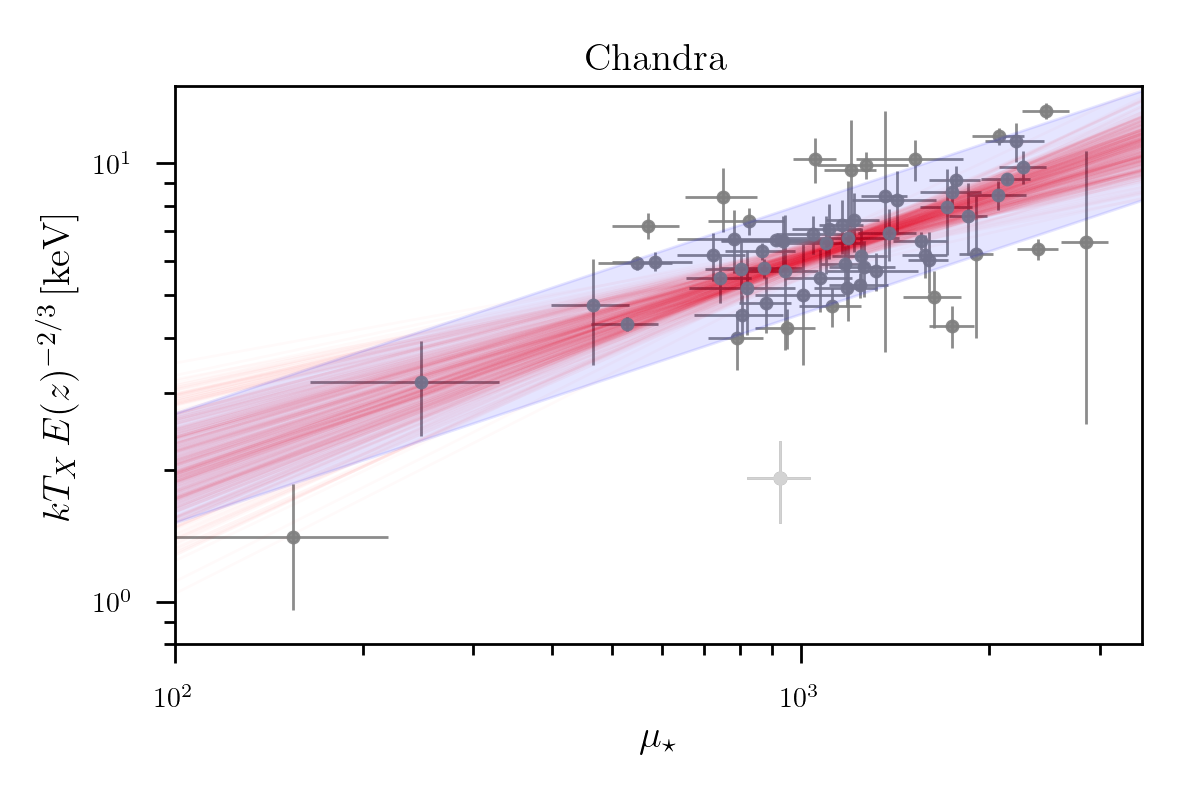}\includegraphics[width=0.45\textwidth]{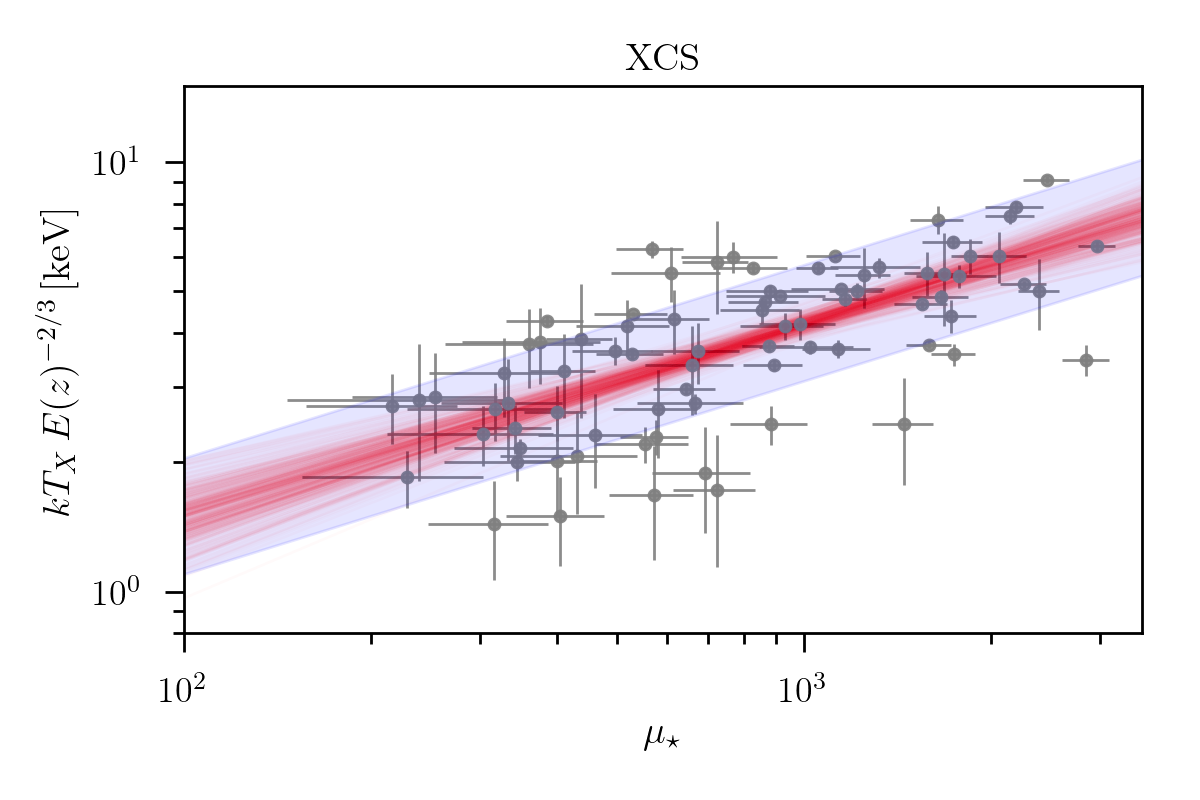}
\includegraphics[width=0.7\textwidth]{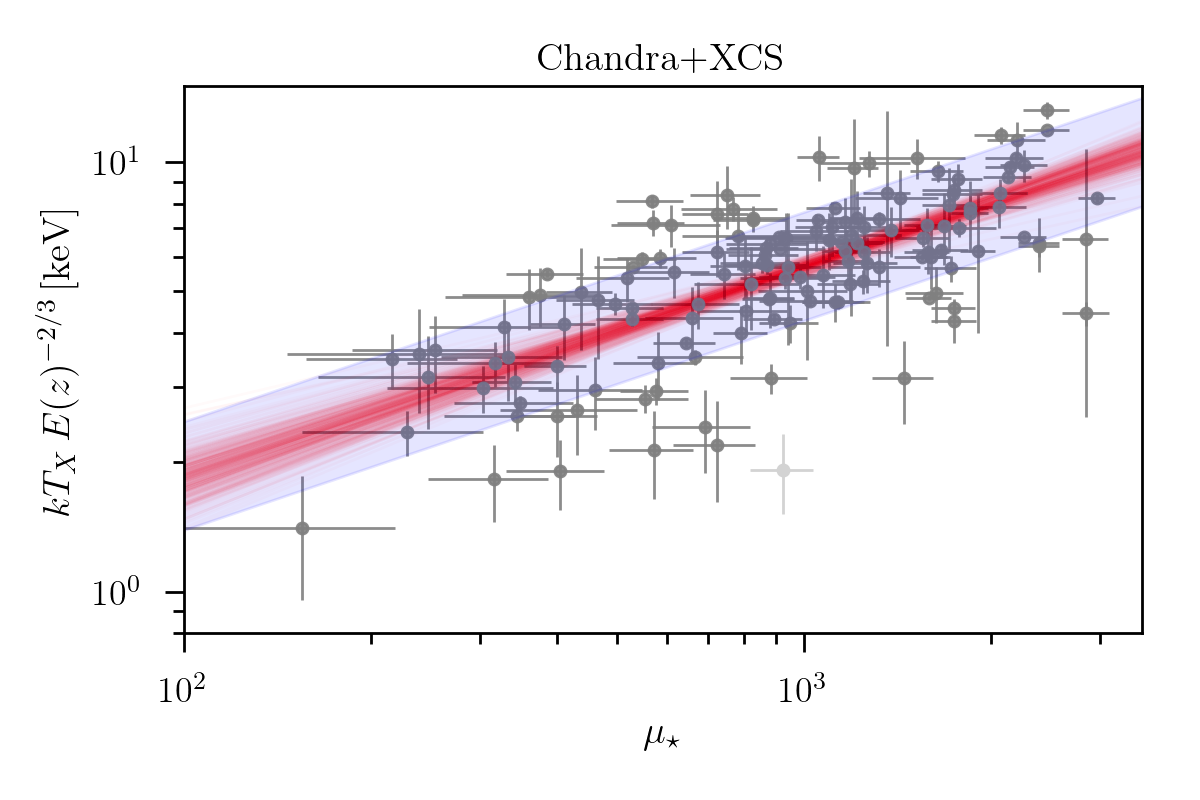}\caption{Bayesian linear regression of $X$-ray temperature and $\mu_\star$ for the \emph{Chandra} and {XMM} samples (top panels) and for the joint sample (bottom panel). The red lines are a random sample from the posterior distribution of slope and intercept, and the blue band represents 1$\sigma$ around the mean value of the intercept plus the intrinsic scatter. The light grey point has been excluded from the regression because it is an outlier with significant deviation from the mean relation ($>3\sigma_{{\rm ln} T|\mu_\star}$).}\label{fig:tx}\end{figure*}

\subsection{X--ray scaling relations}
We first fit the scaling relation presented in Section \ref{sec:xray} for the \emph{XMM} and \emph{Chandra} samples separately. The results of the regression are reported in Table \ref{tab:tx} and shown in Figure \ref{fig:tx}. A few outliers are clearly visible in both samples, particularly in the low--$T_X$ regime. Only one data point (shown in lighter grey in Figure \ref{fig:tx}) has a significant deviation from the mean relation ($>3\sigma_{{\rm ln} T|\mu_\star}$), and it has been excluded from the regression. These outliers tend to have a higher $\mu_\star$ than expected from the mean scaling relation. It is likely that these estimates are affected by the presence of structure along the line of sight, which boosts the mass proxy value. A similar behaviour is also found in \citet{farahi} for the same clusters for the redMaPPer richness, which is computed through a very different methodology, meaning that the outliers are likely related to the galaxies in the DES catalogue rather than to the method.

{We have tested the dependence of the scaling relation results on the completeness of the cluster catalogue. In fact, the X--ray catalogue is likely to be incomplete at the low--temperature end. \citet{farahi} find that the X--ray matched redMaPPer catalogue is $\sim 50$ per cent complete at $\lambda \sim 100$. This corresponds to $\mu_\star\sim 1000$ based on the stellar mass--richness relation found in \cite{palmese19}. We find that, cutting our cluster sample at $\mu_\star>\mu_\star^{\rm cut}=1000$ or higher values, provides scaling relation constraints which are less stringent than those reported in Table \ref{tab:tx}, but still consistent within $1\sigma$.}


The weak lensing mass--$\mu_\star$ relation studied in \citet{maria} shows a steeper slope ($1.74\pm 0.62$ at $0.1<z<0.33$ for SDSS redMaPPer clusters) than the analysis presented here. We believe that the correlation of stellar mass with total cluster mass is higher than with the X--ray temperatures because the X--ray measurement only probes the inner part of the cluster gravitational potential (within $R_{500c}$ and $R_{2500c}$ for the \emph{XMM} and \emph{Chandra} data respectively), while the weak lensing probes larger radii, thus correlating better with the total stellar mass content.


\subsection{Intrinsic temperature scatter}\label{sec:tscatter}

We find an intrinsic scatter in temperature at fixed $\mu_\star$ of $\sigma_{{\rm ln} T_X|\mu_\star}=0.277^{+0.026}_{-0.029}$ for the \emph{XCS} sample. This value is {consistent with the value found for the redMaPPer optical--richness ($0.289\pm 0.025$) in \citet{farahi} within $1\sigma$}.\footnote{$\sigma$ here refers to the error on the scatter.}
The scatter on ${\rm ln}~ T_X$ from the \emph{Chandra} sample is even lower ($0.229^{+0.025}_{-0.027}$) and it is also consistent with the redMaPPer richness estimate ($0.260\pm0.032$) within $1\sigma$. The joint scatter for the two samples is $0.266^{+0.019}_{-0.020}$. {The redMaPPer richness is an optimized \emph{count} observable, and the stellar mass observable has a consistent scatter. We expect $\mu_\star$ to be affected by projection effects similarly to $\lambda$ (as described in \citealt{projection}) or somewhat worse (if the photo--$z$'s do not perform well). It should thus have similar or smaller scatter on the basis of individual halos than is possible from counts alone.}



We perform a number of tests to understand if the membership probabilities are taken into account in an optimal way. We find that including the blue cloud galaxies does not bring a significant increase in the scatter: the inclusion of the second term in the right-hand side of Eq. (\ref{eq:pc}), compared to having solely the red sequence term or redMaPPer members, brings an additional scatter which is an order of magnitude {smaller than the error on that scatter}. This is consistent with what we would expect for this sample, as it has been matched to a red-sequence cluster finder. \citet{extrinsicscatter} found that the blue galaxies significantly increase the scatter of their sample, but the fact that this is not true in our case allows us to keep this contribution which may become relevant at low richness and high redshift regimes, which should be tested using a non-red sequence based cluster finder and matched against other mass observables. It is beyond the scope of this work to test this hypothesis. \citet{extrinsicscatter} also show that differences between the true and predicted scatter of the mass proxy--mass relation are irrelevant for a DES-like survey as long as these differences are about 5\% or less (i.e., $\Delta \sigma < 0.05$), which further supports our choice.

{We also test the effect of the probability $p_R$ in Eq. (12) on $\mu_\star$. We find that the inclusion of this terms makes the mass proxy robust against an arbitrary radial cut between $1$ and $4$ Mpc}: in fact, the intrinsic scatter of the {temperature-$\mu_\star$ relation is insensitive to this choice.} On the other hand, we tested the use of the red galaxies only without including the radial probability contribution. In this case, we find similar trends to previous work (e.g. \citealt{andreon}): optimal choices for the aperture {\em do} exist when no membership probability is considered. {We find that the scatter can decrease with increasing radius, by a factor of up to $\sim 15\%$ with respect to considering the central galaxy only. The maximum improvement is found within a radius typically $<1.5$ Mpc, and outside that radius mostly background/foreground galaxies are added, and the scatter increases with increasing radius.}

We tested the inclusion of colour probabilities $p_c$ in the full membership probabilities by modifying Eq. (\ref{eq:pmem}) into $p_{\rm m} = p_R p_z p_c$. We also tried to combine the colour membership probabilities from different colours in different redshift ranges. This is justified by the fact that most of the colour information in a galaxy SED is contained in the 4000 \AA~ break, that shifts between the bands with redshift. We therefore use $g-r$ for the range $z<0.35$ and $r-i$ in $z>0.35$. We find that these tests did not have a significant impact on the mean scaling relation fit and intrinsic scatter, so it is reasonable to include the simpler version of the full probabilities as given in Eq. (\ref{eq:pmem}).

The fact that our scaling relation scatter and slope are insensitive to the choices made in these tests shows that the membership probabilities are robust and that cluster size and colours (that enter in $p_{\rm m}$ through the redshift probability estimation) are taken into account well.

In order to test the redshift dependence of the scaling relation, we split the joint cluster sample into three subsamples ($z<0.3$, $0.3<z<0.5$ and $z>0.5$) and perform the same linear regression presented above over the whole redshift range. Slope, intercept and scatter are all consistent over the different redshift bins. We conclude that we find no evidence for a redshift evolution of the scaling relation out to $z<0.7$. We believe that this result is related to the fact that the stellar mass content of galaxy clusters is mostly formed before redshift $\sim 1$, and it is consistent with other results (e.g. \citealt{chiu17}) and simulations (\citealt{2018MNRAS.478.2618F}).

\subsection{Total mass scatter}

\begin{figure}
\includegraphics[width=0.45\textwidth]{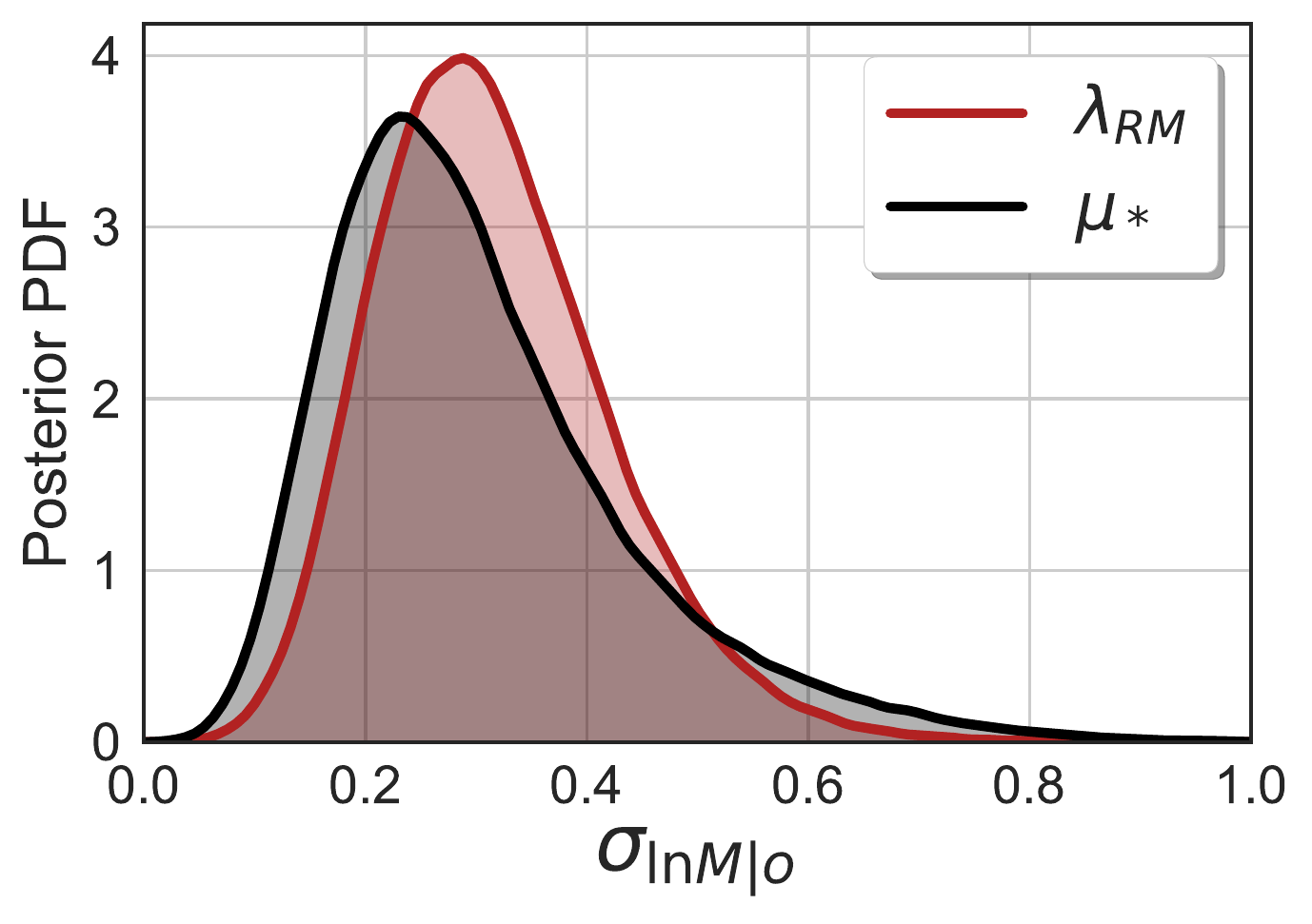}\caption{{Posterior distribution of the scatter in total mass at fixed mass observable $o$ ($\mu_\star$ in black and $\lambda$ in red) for the joint cluster sample. In this work we find: $\sigma_{\ln M|\mu_\star} = 0.26 ^{+ 0.15}_{ - 0.10}$.}}\label{fig:sigma}\end{figure}

\begin{figure}
\includegraphics[width=0.45\textwidth]{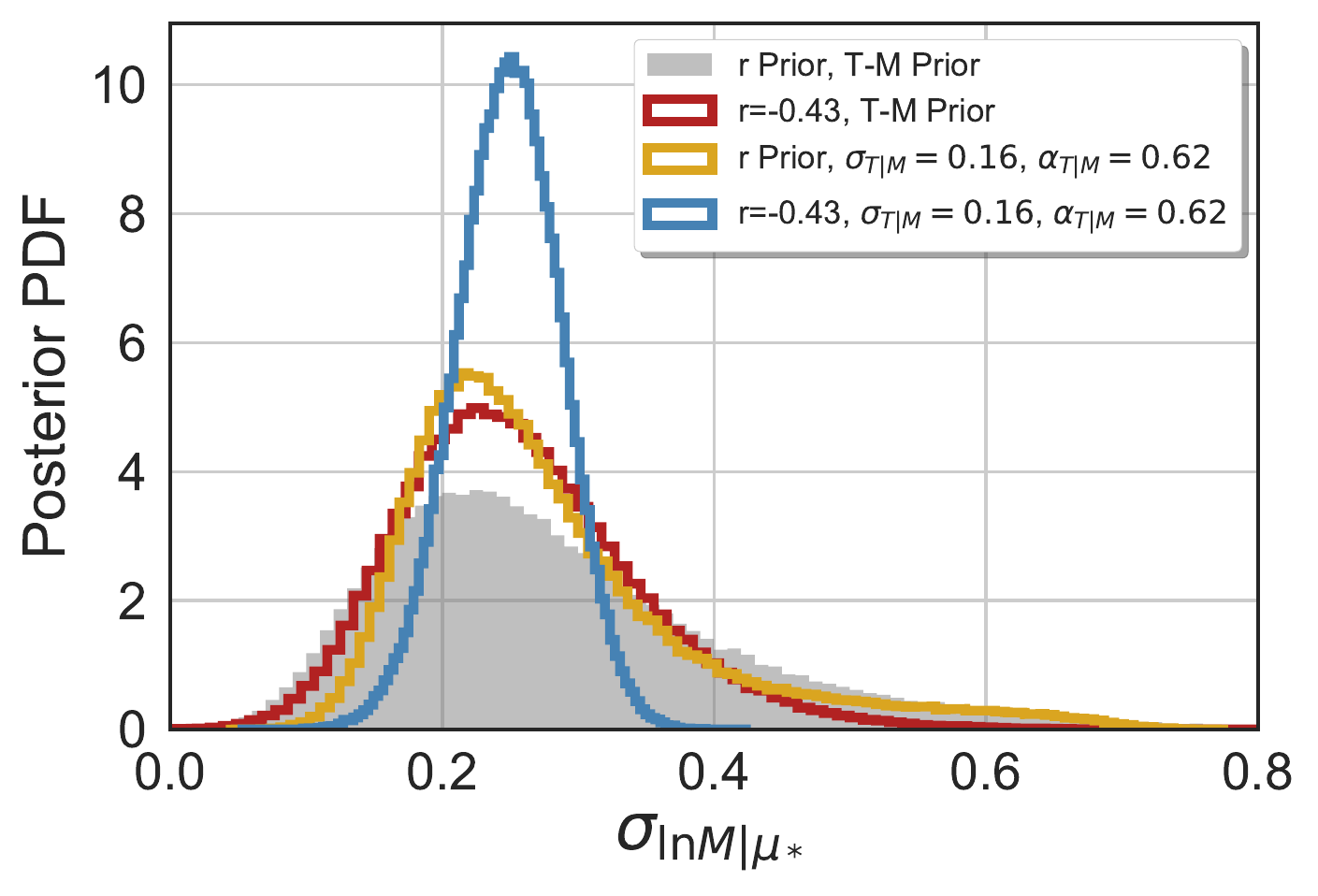}
\caption{Posterior distribution of the scatter in total mass at fixed $\mu_\star$ for the joint cluster sample, showing the impact of different uncertainties on the final PDF. The grey region corresponds to the grey PDF in Figure \ref{fig:sigma}. The other lines show the same pdf, when some of the parameters in the right--hand--side of Eq. (\ref{eq:varmulambda}) are fixed to a known value, and the others are allowed to vary. The red line assumes that the correlation coefficient is fixed at $r_{\mu_\star T}=-0.43$, while the yellow line is computed by fixing the slope and scatter in the $T-M$ relation. The blue line fixes all those quantities, except for the $T-\mu_\star$ scatter derived in this work.}\label{fig:sigma-talk}\end{figure}

\begin{table}\centering
\begin{tabular}{ccc}
\hline
\hline
Parameter & Value & Sample\\
\hline
$\sigma_{\ln T| M}$& $0.16\pm 0.02$ & Weighing the Giants\\
$\alpha_{T| M}$& $0.62\pm 0.04$ & Weighing the Giants\\
$r_{\mu_\star T}$& $-0.43^{+0.49}_{-0.35}$ & LoCuSS\\
\hline
$\sigma_{\ln M|\mu_\star}$ & $0.19 ^{+ 0.15}_{ - 0.09}$ & \emph{Chandra}\\
$\sigma_{\ln M|\mu_\star}$& $0.28 ^{+ 0.16}_{ - 0.11}$ &\emph{XMM}\\
$\sigma_{\ln M|\mu_\star}$& $0.26 ^{+ 0.15}_{ - 0.10}$& \emph{Chandra}+\emph{XMM}\\
\end{tabular}\caption{Parameters used to estimate the mass scatter in Eq. (\ref{eq:varmulambda}) (upper part of the table), and results from this work (lower table).}\label{tab:scatter}
\end{table}

In order to estimate the scatter in mass at fixed $\mu_\star$ presented in Eq. (\ref{eq:varmulambda}) we need to have an estimation of the correlation coefficient $r_{\mu_\star T}$, the scatter in temperature at fixed total mass and the scatter in mass at fixed temperature. The correlation coefficient of pairs of nine observables is estimated by {\citet{Farahi:2019corr}} by employing multi-wavelength analysis of 41 X--ray selected cluster from the LoCuSS clusters sample \citep{mulroy2019}\footnote{{The full posterior chains are publicly available in a figshare repository, \url{https://doi.org/10.6084/m9.figshare.8001218}}}. We employ their $L_K$--$T_X$ correlation coefficient which serves as a good approximation of the stellar mass--temperature correlation in our sample. Their posterior estimate of the correlation coefficient is $r_{\mu_\star T}=-0.43^{+0.49}_{-0.35}$. {Note that this posterior has support over the whole range of possible values ($[-1,1]$), hence the very broad 1$\sigma$ interval reported here. Since this posterior only excludes the extreme tails which are not physically realistic, we believe that it is reasonable to use this correlation coefficient estimate in our analysis, despite the fact that the details in the derivation of temperatures and the redshift range (extending only out to $z\sim 0.3$) in \citet{mulroy2019} may differ from what described here.} The additional parameters needed in Eq. (\ref{eq:varmulambda}), namely $\sigma_{\ln T | M}$ and $\sigma_{\ln M|T}$, are taken from the recent constraints on the scaling relation between the temperature and total mass from weak lensing for the Weighing the Giants program (\citealt{Mantz:2015:wtgI,Mantz:2016WtG-V}). Their posterior {constraints} read $\sigma_{\ln T | M}=0.16 \pm 0.02$ and $\alpha_{T|M} = 0.62 \pm 0.4$. We employ $\sigma_{\ln M|T} = \sigma_{\ln T| M} / \alpha_{T| M}$ to get an {estimate} of mass scatter at fixed X--ray temperature. The posterior distribution of $\sigma_{\ln M|\mu_\star}$ is then obtained by Monte Carlo sampling the right-hand-side of Eq. (\ref{eq:varmulambda}). 

The result for the joint X--ray sample is shown in Figure \ref{fig:sigma}. For the \emph{Chandra} and \emph{XMM} samples we find $\sigma_{\ln M|\mu_\star} = 0.19 ^{+ 0.15}_{ - 0.09}$ and $0.28 ^{+ 0.16}_{ - 0.11}$ respectively, while from the joint analysis $\sigma_{\ln M|\mu_\star} = 0.26 ^{+ 0.15}_{ - 0.10}$. A summary of the parameters used and of these results is reported in Table \ref{tab:scatter}. The errors on the scatter are dominated by the uncertainty on the external parameters described above. In fact, Figure \ref{fig:sigma-talk} shows that the marginalization over the temperature--mass relation (red line) and over the correlation coefficient (yellow line) have a very similar impact on the final posterior estimate, and they dominate the final uncertainty compared to a marginalization over $\sigma_{\ln T|\mu_\star}$ only. 

The scatter found here is consistent with what \citet{farahi} finds for the redMaPPer richness. The slight difference is mostly driven by the lower scatter in temperature at fixed $\mu_\star$ described in Section \ref{sec:tscatter} for the \emph{Chandra} sample. We believe that there is room for further improvement through a more precise estimation of the correlation coefficient, which is driving the more extended tail in Figure \ref{fig:sigma} for $\mu_\star$ compared to the richness case. 

{We note that generalization of these mass scatter findings to clusters with low temperature, $\mu_\star$ or mass should be taken with caution.
Even if our sample is optically selected, we still require a detection in X-ray, with some SNR cut. As a result, both our X-ray sample and the \citet{Mantz:2016WtG-V} sample cover the high--temperature end of the scaling relations. Moreover, \citet{Mantz:2016WtG-V} exclude the cores when measuring X-ray temperatures: it is thus possible that their scatter is smaller than what is applicable to this work. In order to test the effect of different $T-M$ scatter and scaling relations on our results, we have assumed different scaling relation measurements to derive the mass scatter at fixed $\mu_\star$. In particular, we used the results from \citet{2013ApJ...767..116M} and \citet{2016A&A...592A...4L}, who find a lower and larger scatter than \citet{Mantz:2016WtG-V}, respectively. 
Since the uncertainties on the scaling relation from these works are larger than those from \citet{Mantz:2016WtG-V}, the derived mass scatter constraints are weaker, but all consistent with each other. Regardless of the specific value measured for $\sigma_{\ln M|\mu_\star}$, which depends on the analysis choices, one of the main results of this work still holds: the scatter found for this newly established proxy is comparable with the $\lambda$ measurement. }


Remarkably, the scatter found in this work is also consistent with what \citet{2018MNRAS.478.2618F} find {for the total cluster mass at fixed stellar mass} using BAHAMAS and MACSIS simulations, and a similar approach based on \citet{E14}. We assume that $\mu_\star$ probes well the stellar mass content of clusters from BAHAMAS and MACSIS. For the cluster mass probed here ($M_{200c}\sim 10^{14}M_\odot$) we can derive the stellar--mass--conditioned scatter in total mass from their results through: $\sigma_{\ln M|\mu_\star}\simeq \sigma_{\mu_\star|\ln M}/\alpha_{\mu_\star}\simeq 0.22$.

\section{Conclusions}\label{sec:conclusion}

In this work we present a stellar--mass--based mass proxy, $\mu_\star$, and its application to DES Y1 redMaPPer clusters using DES Y3 photometry. In particular, we present a code that uses Bayesian Model Averaging to compute galaxy stellar masses and other galaxy properties. The outputs of this code, along with galaxy membership probabilities presented in a companion paper, are used to estimate our mass proxy. We match Y1 redMaPPer clusters to archival \emph{XMM} and \emph{Chandra} data in order to study the scaling relation of $\mu_\star$ with X--ray temperature. 
Assuming that the scatter in temperature around the mean of the scaling relation at a given $\mu_\star$ is lognormal, and that the temperature scales linearly with $\mu_\star$ in lognormal space, we find that our mass proxy correlates well with the X--ray temperature, with a low intrinsic scatter. Namely, we find that the slope of the scaling relation is $\alpha_{T|\mu_\star}=0.488^{+0.043}_{-0.043}$ and the scatter is $0.266^{+0.019}_{-0.020}$ for the joint \emph{XMM} and \emph{Chandra} cluster sample. This scatter is consistent with what {is} found in a simulation study by \citet{2018MNRAS.478.2618F}. The scaling relation parameters do not show evidence for a deviation from self--similar evolution.

Constraints on the scaling relation between the temperature and total mass from the Weighing the Giants program by \citet{Mantz:2016WtG-V}, along with the luminosity--temperature correlation coefficient estimated by \citet{mulroy2019} on the LoCuSS sample, are then used to derive the expected scatter on halo mass at fixed $\mu_\star$.  We find $\sigma_{{\rm ln} M|\mu_\star}=0.26 ^{+ 0.15}_{ - 0.10}$ for the joint \emph{XMM} and \emph{Chandra} sample. The large uncertainty on this parameter is driven by the marginalisation over the temperature--mass relation parameters and over the correlation coefficient. Consistent values are also found with the same analysis for the well--established redMaPPer mass proxy $\lambda$, showing that $\mu_\star$ is also a potential mass observable to be employed in cosmological analyses with cluster abundances. As such, the mass scatter constrained in this work could serve as a prior on the scatter assumed in the mass observable--mass relation in a cosmological analysis of DES Y1 redMaPPer clusters employing weak lensing measurements. 

It is worth noting that using the stellar mass content of galaxy clusters as mass proxies is empirically and physically motivated, and that measurements of $\mu_\star$ also allow straightforward constraints on the stellar--to--halo connection in clusters. In other words, it allows us to better understand how galaxies evolve in clusters through estimates of their stellar content, while providing a useful tool for cosmological analyses. 
{The other motivation for using this mass proxy rather than galaxy counts is that cluster stellar mass has the potential to be less sensitive to projection effects, one of the main sources of systematics for cluster finders when using photometric data.}

Overall, our results show that $\mu_\star$ is a promising low--scatter mass proxy, which can be used as an alternative to $\lambda$, or in conjunction following the formalism by \citet{E14}, for cosmological and astrophysical analyses with redMaPPer clusters. Future work will also include the development of a new version of the Voronoi--Tessellation cluster finder (\citealt{vt}), that integrates this mass proxy into the pipeline (Burgad et al., in prep.).

\section*{Acknowledgments}
A. Palmese acknowledges the UCL PhD studentship and the URA Visiting scholar award. A. Farahi is supported by a McWilliams Postdoctoral Fellowship. O. Lahav acknowledges support from a European Research Council Advanced Grant FP7/291329. S. Bhargava acknowledges support from the UK Science and Technology Facilities Council via Research Training Grant ST/N504452/1. \textsc{TOPCAT} \citep{topcat} has been extensively used in this work.\\
Funding for the DES Projects has been provided by the U.S. Department of Energy, the U.S. National Science Foundation, the Ministry of Science and Education of Spain, 
the Science and Technology Facilities Council of the United Kingdom, the Higher Education Funding Council for England, the National Center for Supercomputing 
Applications at the University of Illinois at Urbana-Champaign, the Kavli Institute of Cosmological Physics at the University of Chicago, 
the Center for Cosmology and Astro-Particle Physics at the Ohio State University,
the Mitchell Institute for Fundamental Physics and Astronomy at Texas A\&M University, Financiadora de Estudos e Projetos, 
Funda{\c c}{\~a}o Carlos Chagas Filho de Amparo {\`a} Pesquisa do Estado do Rio de Janeiro, Conselho Nacional de Desenvolvimento Cient{\'i}fico e Tecnol{\'o}gico and 
the Minist{\'e}rio da Ci{\^e}ncia, Tecnologia e Inova{\c c}{\~a}o, the Deutsche Forschungsgemeinschaft and the Collaborating Institutions in the Dark Energy Survey. 

The Collaborating Institutions are Argonne National Laboratory, the University of California at Santa Cruz, the University of Cambridge, Centro de Investigaciones Energ{\'e}ticas, 
Medioambientales y Tecnol{\'o}gicas-Madrid, the University of Chicago, University College London, the DES-Brazil Consortium, the University of Edinburgh, 
the Eidgen{\"o}ssische Technische Hochschule (ETH) Z{\"u}rich, 
Fermi National Accelerator Laboratory, the University of Illinois at Urbana-Champaign, the Institut de Ci{\`e}ncies de l'Espai (IEEC/CSIC), 
the Institut de F{\'i}sica d'Altes Energies, Lawrence Berkeley National Laboratory, the Ludwig-Maximilians Universit{\"a}t M{\"u}nchen and the associated Excellence Cluster Universe, 
the University of Michigan, the National Optical Astronomy Observatory, the University of Nottingham, The Ohio State University, the University of Pennsylvania, the University of Portsmouth, 
SLAC National Accelerator Laboratory, Stanford University, the University of Sussex, and Texas A\&M University.

The DES data management system is supported by the National Science Foundation under Grant Number AST-1138766.
The DES participants from Spanish institutions are partially supported by MINECO under grants AYA2012-39559, ESP2013-48274, FPA2013-47986, and Centro de Excelencia Severo Ochoa SEV-2012-0234.
Research leading to these results has received funding from the European Research Council under the European Unions Seventh Framework Programme (FP7/2007-2013) including ERC grant agreements 
 240672, 291329, and 306478.
 
\bibliographystyle{mn2e}
\bibliography{bib}

\begin{thebibliography}{86}
\expandafter\ifx\csname natexlab\endcsname\relax\def\natexlab#1{#1}\fi

\bibitem[{{Abbott} {et~al}\mbox{.}(2018){Abbott}, {Abdalla}, {Allam}, {Amara},
  {Annis}, {Asorey}, {Avila}, {Ballester}, {Banerji}, {Barkhouse}, {Baruah},
  {Baumer}, {Bechtol}, {Becker}, {Benoit-L{\'e}vy}, {Bernstein}, {Bertin},
  {Blazek}, {Bocquet}, {Brooks}, {Brout}, {Buckley-Geer}, {Burke}, {Busti},
  {Campisano}, {Cardiel-Sas}, {Carnero Rosell}, {Carrasco Kind}, {Carretero},
  {Castander}, {Cawthon}, {Chang}, {Chen}, {Conselice}, {Costa}, {Crocce},
  {Cunha}, {D{\textquoteright}Andrea}, {da Costa}, {Das}, {Daues}, {Davis},
  {Davis}, {De Vicente}, {DePoy}, {DeRose}, {Desai}, {Diehl}, {Dietrich},
  {Dodelson}, {Doel}, {Drlica-Wagner}, {Eifler}, {Elliott}, {Evrard}, {Farahi},
  {Fausti Neto}, {Fernand ez}, {Finley}, {Flaugher}, {Foley}, {Fosalba},
  {Friedel}, {Frieman}, {Garc{\'\i}a-Bellido}, {Gaztanaga}, {Gerdes},
  {Giannantonio}, {Gill}, {Glazebrook}, {Goldstein}, {Gower}, {Gruen},
  {Gruendl}, {Gschwend}, {Gupta}, {Gutierrez}, {Hamilton}, {Hartley}, {Hinton},
  {Hislop}, {Hollowood}, {Honscheid}, {Hoyle}, {Huterer}, {Jain}, {James},
  {Jeltema}, {Johnson}, {Johnson}, {Kacprzak}, {Kent}, {Khullar}, {Klein},
  {Kovacs}, {Koziol}, {Krause}, {Kremin}, {Kron}, {Kuehn}, {Kuhlmann},
  {Kuropatkin}, {Lahav}, {Lasker}, {Li}, {Li}, {Liddle}, {Lima}, {Lin},
  {L{\'o}pez-Reyes}, {MacCrann}, {Maia}, {Maloney}, {Manera}, {March},
  {Marriner}, {Marshall}, {Martini}, {McClintock}, {McKay}, {McMahon},
  {Melchior}, {Menanteau}, {Miller}, {Miquel}, {Mohr}, {Morganson}, {Mould},
  {Neilsen}, {Nichol}, {Nogueira}, {Nord}, {Nugent}, {Nunes}, {Ogand o}, {Old},
  {Pace}, {Palmese}, {Paz-Chinch{\'o}n}, {Peiris}, {Percival}, {Petravick},
  {Plazas}, {Poh}, {Pond}, {Porredon}, {Pujol}, {Refregier}, {Reil}, {Ricker},
  {Rollins}, {Romer}, {Roodman}, {Rooney}, {Ross}, {Rykoff}, {Sako}, {Sanchez},
  {Sanchez}, {Santiago}, {Saro}, {Scarpine}, {Scolnic}, {Serrano},
  {Sevilla-Noarbe}, {Sheldon}, {Shipp}, {Silveira}, {Smith}, {Smith}, {Smith},
  {Soares-Santos}, {Sobreira}, {Song}, {Stebbins}, {Suchyta}, {Sullivan},
  {Swanson}, {Tarle}, {Thaler}, {Thomas}, {Thomas}, {Troxel}, {Tucker},
  {Vikram}, {Vivas}, {Walker}, {Wechsler}, {Weller}, {Wester}, {Wolf}, {Wu},
  {Yanny}, {Zenteno}, {Zhang}, {Zuntz}, {DES Collaboration}, {Juneau},
  {Fitzpatrick}, {Nikutta}, {Nidever}, {Olsen}, {Scott}, \& {Data Lab}}]{dr1}
{Abbott} T.~M.~C. {et~al.}, 2018, The Astrophysical Journal Supplement Series,
  239, 18

\bibitem[{{Andreon}(2012)}]{andreon12}
{Andreon} S., 2012, \aap, 548, A83

\bibitem[{{Andreon}(2015)}]{andreon}
{Andreon} S., 2015, \aap, 582, A100

\bibitem[{{Arnaud}(1996)}]{Arnaud96}
{Arnaud} K.~A., 1996, in Astronomical Society of the Pacific Conference Series,
  Vol. 101, Astronomical Data Analysis Software and Systems V, {Jacoby} G.~H.,
  {Barnes} J., eds., p.~17

\bibitem[{Arnaud, Pointecouteau \& Pratt(2005)Arnaud, Pointecouteau, \&
  Pratt}]{Arnaud2005}
Arnaud M., Pointecouteau E., Pratt G.~W., 2005, Astronomy and Astrophysics,
  441, 893

\bibitem[{{Ben{\'{\i}}tez}(2000)}]{bpz}
{Ben{\'{\i}}tez} N., 2000, \apj, 536, 571

\bibitem[{{Bertin} \& {Arnouts}(1996)}]{sextractor}
{Bertin} E., {Arnouts} S., 1996, Astronomy and Astrophysics Supplement, 117,
  393

\bibitem[{{Blanton} \& {Roweis}(2007)}]{blanton}
{Blanton} M.~R., {Roweis} S., 2007, \aj, 133, 734

\bibitem[{{Bradshaw} {et~al}\mbox{.}(2019){Bradshaw}, {Leauthaud}, {Hearin},
  {Huang}, \& {Behroozi}}]{Bradshaw}
{Bradshaw} C., {Leauthaud} A., {Hearin} A., {Huang} S., {Behroozi} P., 2019,
  arXiv e-prints, arXiv:1905.09353

\bibitem[{{Bruzual} \& {Charlot}(2003)}]{bc03}
{Bruzual} G., {Charlot} S., 2003, \mnras, 344, 1000

\bibitem[{{Bryan} \& {Norman}(1998)}]{BryanNorman:1998}
{Bryan} G.~L., {Norman} M.~L., 1998, \apj, 495, 80

\bibitem[{{Butcher} \& {Oemler}(1978)}]{butcher1}
{Butcher} H., {Oemler}, Jr. A., 1978, \apj, 226, 559

\bibitem[{{Butcher} \& {Oemler}(1984)}]{butcher2}
{Butcher} H., {Oemler}, Jr. A., 1984, \apj, 285, 426

\bibitem[{{Capozzi} {et~al}\mbox{.}(2017){Capozzi}, {Etherington}, {Thomas},
  {Maraston}, {Rykoff}, {Sevilla-Noarbe}, {Bechtol}, {Carrasco Kind},
  {Drlica-Wagner}, {Pforr}, {Gschwend}, {Carnero Rosell}, {Pellegrini}, {Maia},
  {da Costa}, {Benoit-L{\'e}vy}, {Swanson}, {Wechsler}, {Banerji}, {Papovich},
  {Morice-Atkinson}, {Abdalla}, {Brooks}, {Carretero}, {Cunha}, {D'Andrea},
  {Desai}, {Diehl}, {Evrards}, {Flaugher}, {Fosalba}, {Frieman},
  {Garc{\'{\i}}a-Bellido}, {Gaztanaga}, {Gerdes}, {Gruen}, {Gruendl},
  {Gutierrez}, {Hartley}, {James}, {Jeltema}, {Kuehn}, {Kuhlmann},
  {Kuropatkin}, {Lahav}, {Lima}, {Marshall}, {Martini}, {Menanteau}, {Miquel},
  {Nord}, {Ogando}, {Plazas Malag{\`o}n}, {Romer}, {Sanchez}, {Scarpine},
  {Schindler}, {Schubnell}, {Smith}, {Soares-Santos}, {Sobreira}, {Suchyta}, \&
  {Tarle}}]{capozzi}
{Capozzi} D. {et~al.}, 2017, ArXiv e-prints

\bibitem[{{Chiu} {et~al}\mbox{.}(2018){Chiu}, {Mohr}, {McDonald}, {Bocquet},
  {Desai}, {Klein}, {Israel}, {Ashby}, {Stanford}, {Benson}, {Brodwin},
  {Abbott}, {Abdalla}, {Allam}, {Annis}, {Bayliss}, {Benoit-L{\'e}vy},
  {Bertin}, {Bleem}, {Brooks}, {Buckley-Geer}, {Bulbul}, {Capasso},
  {Carlstrom}, {Rosell}, {Carretero}, {Castander}, {Cunha}, {D'Andrea}, {da
  Costa}, {Davis}, {Diehl}, {Dietrich}, {Doel}, {Drlica-Wagner}, {Eifler},
  {Evrard}, {Flaugher}, {Garc{\'{\i}}a-Bellido}, {Garmire}, {Gaztanaga},
  {Gerdes}, {Gonzalez}, {Gruen}, {Gruendl}, {Gschwend}, {Gupta}, {Gutierrez},
  {Hlavacek-L}, {Honscheid}, {James}, {Jeltema}, {Kraft}, {Krause}, {Kuehn},
  {Kuhlmann}, {Kuropatkin}, {Lahav}, {Lima}, {Maia}, {Marshall}, {Melchior},
  {Menanteau}, {Miquel}, {Murray}, {Nord}, {Ogando}, {Plazas}, {Rapetti},
  {Reichardt}, {Romer}, {Roodman}, {Sanchez}, {Saro}, {Scarpine}, {Schindler},
  {Schubnell}, {Sharon}, {Smith}, {Smith}, {Soares-Santos}, {Sobreira},
  {Stalder}, {Stern}, {Strazzullo}, {Suchyta}, {Swanson}, {Tarle}, {Vikram},
  {Walker}, {Weller}, \& {Zhang}}]{chiu17}
{Chiu} I. {et~al.}, 2018, \mnras, 478, 3072

\bibitem[{{Coleman}, {Wu} \& {Weedman}(1980){Coleman}, {Wu}, \&
  {Weedman}}]{Coleman}
{Coleman} G.~D., {Wu} C.-C., {Weedman} D.~W., 1980, \apjs, 43, 393

\bibitem[{{Conroy} \& {Gunn}(2010)}]{fsps}
{Conroy} C., {Gunn} J.~E., 2010, \apj, 712, 833

\bibitem[{{Cool} {et~al}\mbox{.}(2013){Cool}, {Moustakas}, {Blanton}, {Burles},
  {Coil}, {Eisenstein}, {Wong}, {Zhu}, {Aird}, {Bernstein}, {Bolton}, {Hogg},
  \& {Mendez}}]{Cool}
{Cool} R.~J. {et~al.}, 2013, \apj, 767, 118

\bibitem[{{Costanzi} {et~al}\mbox{.}(2019){Costanzi}, {Rozo}, {Rykoff},
  {Farahi}, {Jeltema}, {Evrard}, {Mantz}, {Gruen}, {Mandelbaum}, {DeRose},
  {McClintock}, {Varga}, {Zhang}, {Weller}, {Wechsler}, \&
  {Aguena}}]{projection}
{Costanzi} M. {et~al.}, 2019, \mnras, 482, 490

\bibitem[{{De Lucia} \& {Blaizot}(2007)}]{delucia}
{De Lucia} G., {Blaizot} J., 2007, \mnras, 375, 2

\bibitem[{{Desai} {et~al}\mbox{.}(2012){Desai}, {Armstrong}, {Mohr}, {Semler},
  {Liu}, {Bertin}, {Allam}, {Barkhouse}, {Bazin}, {Buckley-Geer}, {Cooper},
  {Hansen}, {High}, {Lin}, {Lin}, {Ngeow}, {Rest}, {Song}, {Tucker}, \&
  {Zenteno}}]{desai}
{Desai} S. {et~al.}, 2012, \apj, 757, 83

\bibitem[{{Diehl} {et~al}\mbox{.}(2014){Diehl}, {Abbott}, {Annis}, {Armstrong},
  {Baruah}, {Bermeo}, {Bernstein}, {Beynon}, {Bruderer}, {Buckley-Geer},
  {Campbell}, {Capozzi}, {Carter}, {Casas}, {Clerkin}, {Covarrubias}, {Cuhna},
  {D'Andrea}, {da Costa}, {Das}, {DePoy}, {Dietrich}, {Drlica-Wagner},
  {Elliott}, {Eifler}, {Estrada}, {Etherington}, {Flaugher}, {Frieman}, {Fausti
  Neto}, {Gelman}, {Gerdes}, {Gruen}, {Gruendl}, {Hao}, {Head}, {Helsby},
  {Hoffman}, {Honscheid}, {James}, {Johnson}, {Kacprzac}, {Katsaros},
  {Kennedy}, {Kent}, {Kessler}, {Kim}, {Krause}, {Kron}, {Kuhlmann}, {Kunder},
  {Li}, {Lin}, {Maccrann}, {March}, {Marshall}, {Neilsen}, {Nugent}, {Martini},
  {Melchior}, {Menanteau}, {Nichol}, {Nord}, {Ogando}, {Old}, {Papadopoulos},
  {Patton}, {Petravick}, {Plazas}, {Poulton}, {Pujol}, {Reil}, {Rigby},
  {Romer}, {Roodman}, {Rooney}, {Sanchez Alvaro}, {Serrano}, {Sheldon},
  {Smith}, {Smith}, {Soares-Santos}, {Soumagnac}, {Spinka}, {Suchyta},
  {Tucker}, {Walker}, {Wester}, {Wiesner}, {Wilcox}, {Williams}, {Yanny}, \&
  {Zhang}}]{y1}
{Diehl} H.~T. {et~al.}, 2014, in Society of Photo-Optical Instrumentation
  Engineers (SPIE) Conference Series, Vol. 9149, Society of Photo-Optical
  Instrumentation Engineers (SPIE) Conference Series, p. 91490V

\bibitem[{{Donahue} {et~al}\mbox{.}(2002){Donahue}, {Scharf}, {Mack}, {Lee},
  {Postman}, {Rosati}, {Dickinson}, {Voit}, \& {Stocke}}]{Donahue}
{Donahue} M. {et~al.}, 2002, \apj, 569, 689

\bibitem[{{Drlica-Wagner} {et~al}\mbox{.}(2018){Drlica-Wagner},
  {Sevilla-Noarbe}, {Rykoff}, {Gruendl}, {Yanny}, {Tucker}, {Hoyle}, {Carnero
  Rosell}, {Bernstein}, {Bechtol}, {Becker}, {Benoit-L{\'e}vy}, {Bertin},
  {Carrasco Kind}, {Davis}, {de Vicente}, {Diehl}, {Gruen}, {Hartley},
  {Leistedt}, {Li}, {Marshall}, {Neilsen}, {Rau}, {Sheldon}, {Smith}, {Troxel},
  {Wyatt}, {Zhang}, {Abbott}, {Abdalla}, {Allam}, {Banerji}, {Brooks},
  {Buckley-Geer}, {Burke}, {Capozzi}, {Carretero}, {Cunha}, {D'Andrea}, {da
  Costa}, {DePoy}, {Desai}, {Dietrich}, {Doel}, {Evrard}, {Fausti Neto},
  {Flaugher}, {Fosalba}, {Frieman}, {Garc{\'\i}a-Bellido}, {Gerdes},
  {Giannantonio}, {Gschwend}, {Gutierrez}, {Honscheid}, {James}, {Jeltema},
  {Kuehn}, {Kuhlmann}, {Kuropatkin}, {Lahav}, {Lima}, {Lin}, {Maia}, {Martini},
  {McMahon}, {Melchior}, {Menanteau}, {Miquel}, {Nichol}, {Ogand o}, {Plazas},
  {Romer}, {Roodman}, {Sanchez}, {Scarpine}, {Schindler}, {Schubnell}, {Smith},
  {Smith}, {Soares-Santos}, {Sobreira}, {Suchyta}, {Tarle}, {Vikram}, {Walker},
  {Wechsler}, {Zuntz}, \& {DES Collaboration}}]{firstyear}
{Drlica-Wagner} A. {et~al.}, 2018, The Astrophysical Journal Supplement Series,
  235, 33

\bibitem[{{Evrard} {et~al}\mbox{.}(2014){Evrard}, {Arnault}, {Huterer}, \&
  {Farahi}}]{E14}
{Evrard} A.~E., {Arnault} P., {Huterer} D., {Farahi} A., 2014, \mnras, 441,
  3562

\bibitem[{{Fabian} {et~al}\mbox{.}(1994){Fabian}, {Crawford}, {Edge}, \&
  {Mushotzky}}]{Fabian:1994}
{Fabian} A.~C., {Crawford} C.~S., {Edge} A.~C., {Mushotzky} R.~F., 1994,
  \mnras, 267, 779

\bibitem[{{Farahi} {et~al}\mbox{.}(2019{\natexlab{a}}){Farahi}, {Chen},
  {Evrard}, {Hollowood}, {Wilkinson}, {Bhargava}, {Giles}, {Romer}, {Jeltema},
  {Hilton}, {Bermeo}, {Mayers}, {Cervantes}, {Rozo}, {Rykoff}, {Collins},
  {Costanzi}, {Everett}, {Liddle}, {Mann}, {Mantz}, {Rooney}, {Sahlen},
  {Stott}, {Viana}, {Zhang}, {Annis}, {Avila}, {Brooks}, {Buckley-Geer},
  {Burke}, {Carnero Rosell}, {Carrasco Kind}, {Carretero}, {Castander}, {da
  Costa}, {De Vicente}, {Desai}, {Diehl}, {Dietrich}, {Doel}, {Flaugher},
  {Fosalba}, {Frieman}, {Garc{\'\i}a-Bellido}, {Gaztanaga}, {Gerdes}, {Gruen},
  {Gruendl}, {Gschwend}, {Gutierrez}, {Honscheid}, {James}, {Krause}, {Kuehn},
  {Kuropatkin}, {Lima}, {Maia}, {Marshall}, {Melchior}, {Menanteau}, {Miquel},
  {Ogando}, {Plazas}, {Sanchez}, {Scarpine}, {Schubnell}, {Serrano},
  {Sevilla-Noarbe}, {Smith}, {Sobreira}, {Suchyta}, {Swanson}, {Tarle},
  {Thomas}, {Tucker}, {Vikram}, {Walker}, \& {Weller}}]{farahi}
{Farahi} A. {et~al.}, 2019{\natexlab{a}}, \mnras, 2299

\bibitem[{{Farahi} {et~al}\mbox{.}(2018){Farahi}, {Evrard}, {McCarthy},
  {Barnes}, \& {Kay}}]{2018MNRAS.478.2618F}
{Farahi} A., {Evrard} A.~E., {McCarthy} I., {Barnes} D.~J., {Kay} S.~T., 2018,
  \mnras, 478, 2618

\bibitem[{{Farahi} {et~al}\mbox{.}(2019{\natexlab{b}}){Farahi}, {Mulroy},
  {Evrard}, {Smith}, {Finoguenov}, {Bourdin}, {Carlstrom}, {Haines}, {Marrone},
  {Martino}, {Mazzotta}, {O'Donnell}, \& {Okabe}}]{Farahi:2019corr}
{Farahi} A. {et~al.}, 2019{\natexlab{b}}, Nature Communications, 10

\bibitem[{{Flaugher} {et~al}\mbox{.}(2015){Flaugher}, {Diehl}, {Honscheid},
  {Abbott}, {Alvarez}, {Angstadt}, {Annis}, {Antonik}, {Ballester}, {Beaufore},
  {Bernstein}, {Bernstein}, {Bigelow}, {Bonati}, {Boprie}, {Brooks},
  {Buckley-Geer}, {Campa}, {Cardiel-Sas}, {Castander}, {Castilla}, {Cease},
  {Cela-Ruiz}, {Chappa}, {Chi}, {Cooper}, {da Costa}, {Dede}, {Derylo},
  {DePoy}, {de Vicente}, {Doel}, {Drlica-Wagner}, {Eiting}, {Elliott}, {Emes},
  {Estrada}, {Fausti Neto}, {Finley}, {Flores}, {Frieman}, {Gerdes},
  {Gladders}, {Gregory}, {Gutierrez}, {Hao}, {Holland}, {Holm}, {Huffman},
  {Jackson}, {James}, {Jonas}, {Karcher}, {Karliner}, {Kent}, {Kessler},
  {Kozlovsky}, {Kron}, {Kubik}, {Kuehn}, {Kuhlmann}, {Kuk}, {Lahav}, {Lathrop},
  {Lee}, {Levi}, {Lewis}, {Li}, {Mandrichenko}, {Marshall}, {Martinez},
  {Merritt}, {Miquel}, {Mu{\~n}oz}, {Neilsen}, {Nichol}, {Nord}, {Ogando},
  {Olsen}, {Palaio}, {Patton}, {Peoples}, {Plazas}, {Rauch}, {Reil}, {Rheault},
  {Roe}, {Rogers}, {Roodman}, {Sanchez}, {Scarpine}, {Schindler}, {Schmidt},
  {Schmitt}, {Schubnell}, {Schultz}, {Schurter}, {Scott}, {Serrano}, {Shaw},
  {Smith}, {Soares-Santos}, {Stefanik}, {Stuermer}, {Suchyta}, {Sypniewski},
  {Tarle}, {Thaler}, {Tighe}, {Tran}, {Tucker}, {Walker}, {Wang}, {Watson},
  {Weaverdyck}, {Wester}, {Woods}, {Yanny}, \& {The DES
  Collaboration}}]{flaugher}
{Flaugher} B. {et~al.}, 2015, \aj, 150, 150

\bibitem[{{Giles} {et~al}\mbox{.}(2019){Giles}, {Diehl}, {Honscheid}, {Abbott},
  {Alvarez}, {Angstadt}, {Annis}, {Antonik}, {Ballester}, {Beaufore},
  {Bernstein}, {Bernstein}, {Bigelow}, {Bonati}, {Boprie}, {Brooks},
  {Buckley-Geer}, {Campa}, {Cardiel-Sas}, {Castander}, {Castilla}, {Cease},
  {Cela-Ruiz}, {Chappa}, {Chi}, {Cooper}, {da Costa}, {Dede}, {Derylo},
  {DePoy}, {de Vicente}, {Doel}, {Drlica-Wagner}, {Eiting}, {Elliott}, {Emes},
  {Estrada}, {Fausti Neto}, {Finley}, {Flores}, {Frieman}, {Gerdes},
  {Gladders}, {Gregory}, {Gutierrez}, {Hao}, {Holland}, {Holm}, {Huffman},
  {Jackson}, {James}, {Jonas}, {Karcher}, {Karliner}, {Kent}, {Kessler},
  {Kozlovsky}, {Kron}, {Kubik}, {Kuehn}, {Kuhlmann}, {Kuk}, {Lahav}, {Lathrop},
  {Lee}, {Levi}, {Lewis}, {Li}, {Mandrichenko}, {Marshall}, {Martinez},
  {Merritt}, {Miquel}, {Mu{\~n}oz}, {Neilsen}, {Nichol}, {Nord}, {Ogando},
  {Olsen}, {Palaio}, {Patton}, {Peoples}, {Plazas}, {Rauch}, {Reil}, {Rheault},
  {Roe}, {Rogers}, {Roodman}, {Sanchez}, {Scarpine}, {Schindler}, {Schmidt},
  {Schmitt}, {Schubnell}, {Schultz}, {Schurter}, {Scott}, {Serrano}, {Shaw},
  {Smith}, {Soares-Santos}, {Stefanik}, {Stuermer}, {Suchyta}, {Sypniewski},
  {Tarle}, {Thaler}, {Tighe}, {Tran}, {Tucker}, {Walker}, {Wang}, {Watson},
  {Weaverdyck}, {Wester}, {Woods}, {Yanny}, \& {The DES Collaboration}}]{giles}
{Giles} P. {et~al.}, 2019, in prep.

\bibitem[{{Girardi} {et~al}\mbox{.}(2000){Girardi}, {Bressan}, {Bertelli}, \&
  {Chiosi}}]{padova1}
{Girardi} L., {Bressan} A., {Bertelli} G., {Chiosi} C., 2000, \aaps, 141, 371

\bibitem[{Hao {et~al}\mbox{.}(2009)Hao, Koester, Mckay, Rykoff, Rozo, Evrard,
  Annis, Becker, Busha, Gerdes, Johnston, Sheldon, \&
  Wechsler}]{Hao2009PRECISIONMODEL}
Hao J. {et~al.}, 2009, The Astrophysical Journal, 702, 745

\bibitem[{{Hao} {et~al}\mbox{.}(2010){Hao}, {McKay}, {Koester}, {Rykoff},
  {Rozo}, {Annis}, {Wechsler}, {Evrard}, {Siegel}, {Becker}, {Busha}, {Gerdes},
  {Johnston}, \& {Sheldon}}]{hao}
{Hao} J. {et~al.}, 2010, \apjs, 191, 254

\bibitem[{{Hoeting} {et~al}\mbox{.}(1999){Hoeting}, {Madigan}, {Raftery}, \&
  {Volinsky}}]{hoeting}
{Hoeting} J.~A., {Madigan} D., {Raftery} A.~E., {Volinsky} C.~T., 1999,
  Statist. Sci., 14, 382

\bibitem[{{Hollowood} {et~al}\mbox{.}(2019){Hollowood}, {Jeltema}, {Chen},
  {Farahi}, {Evrard}, {Everett}, {Rozo}, {Rykoff}, {Bernstein},
  {Bermeo-Hernandez}, {Eiger}, {Giles}, {Israel}, {Michel}, {Noorali}, {Romer},
  {Rooney}, \& {Splettstoesser}}]{hollowood}
{Hollowood} D.~L. {et~al.}, 2019, \apjs, 244, 22

\bibitem[{{Hoyle} {et~al}\mbox{.}(2018){Hoyle}, {Gruen}, {Bernstein}, {Rau},
  {De Vicente}, {Hartley}, {Gaztanaga}, {DeRose}, {Troxel}, {Davis}, {Alarcon},
  {MacCrann}, {Prat}, {S{\'a}nchez}, {Sheldon}, {Wechsler}, {Asorey}, {Becker},
  {Bonnett}, {Carnero Rosell}, {Carollo}, {Carrasco Kind}, {Castander},
  {Cawthon}, {Chang}, {Childress}, {Davis}, {Drlica-Wagner}, {Gatti},
  {Glazebrook}, {Gschwend}, {Hinton}, {Hoormann}, {Kim}, {King}, {Kuehn},
  {Lewis}, {Lidman}, {Lin}, {Macaulay}, {Maia}, {Martini}, {Mudd},
  {M{\"o}ller}, {Nichol}, {Ogando}, {Rollins}, {Roodman}, {Ross}, {Rozo},
  {Rykoff}, {Samuroff}, {Sevilla-Noarbe}, {Sharp}, {Sommer}, {Tucker}, {Uddin},
  {Varga}, {Vielzeuf}, {Yuan}, {Zhang}, {Abbott}, {Abdalla}, {Allam}, {Annis},
  {Bechtol}, {Benoit-L{\'e}vy}, {Bertin}, {Brooks}, {Buckley-Geer}, {Burke},
  {Busha}, {Capozzi}, {Carretero}, {Crocce}, {D'Andrea}, {da Costa}, {DePoy},
  {Desai}, {Diehl}, {Doel}, {Eifler}, {Estrada}, {Evrard}, {Fernandez},
  {Flaugher}, {Fosalba}, {Frieman}, {Garc{\'{\i}}a-Bellido}, {Gerdes},
  {Giannantonio}, {Goldstein}, {Gruendl}, {Gutierrez}, {Honscheid}, {James},
  {Jarvis}, {Jeltema}, {Johnson}, {Johnson}, {Kirk}, {Krause}, {Kuhlmann},
  {Kuropatkin}, {Lahav}, {Li}, {Lima}, {March}, {Marshall}, {Melchior},
  {Menanteau}, {Miquel}, {Nord}, {O'Neill}, {Plazas}, {Romer}, {Sako},
  {Sanchez}, {Santiago}, {Scarpine}, {Schindler}, {Schubnell}, {Smith},
  {Smith}, {Soares-Santos}, {Sobreira}, {Suchyta}, {Swanson}, {Tarle},
  {Thomas}, {Tucker}, {Vikram}, {Walker}, {Weller}, {Wester}, {Wolf}, {Yanny},
  \& {Zuntz}}]{Hoyle}
{Hoyle} B. {et~al.}, 2018, \mnras, 478, 592

\bibitem[{{Ivezi{\'c}} {et~al}\mbox{.}(2008){Ivezi{\'c}}, {Kahn}, {Tyson},
  {Abel}, {Acosta}, {Allsman}, {Alonso}, {AlSayyad}, {Anderson}, {Andrew}, \&
  et~al.}]{lsst}
{Ivezi{\'c}} {\v Z}. {et~al.}, 2008, arXiv e-prints

\bibitem[{{Jansen} \& {Laine}(1997)}]{xmm}
{Jansen} F.~A., {Laine} R., 1997, in Bulletin of the American Astronomical
  Society, Vol.~29, American Astronomical Society Meeting Abstracts, p. 1365

\bibitem[{{Kaiser}(1991)}]{Kaiser:1991}
{Kaiser} N., 1991, \apj, 383, 104

\bibitem[{{Kelly}(2007)}]{kelly}
{Kelly} B.~C., 2007, \apj, 665, 1489

\bibitem[{{Kinney} {et~al}\mbox{.}(1996){Kinney}, {Calzetti}, {Bohlin},
  {McQuade}, {Storchi-Bergmann}, \& {Schmitt}}]{Kinney}
{Kinney} A.~L., {Calzetti} D., {Bohlin} R.~C., {McQuade} K., {Storchi-Bergmann}
  T., {Schmitt} H.~R., 1996, \apj, 467, 38

\bibitem[{{Koester} {et~al}\mbox{.}(2007){Koester}, {McKay}, {Annis},
  {Wechsler}, {Evrard}, {Bleem}, {Becker}, {Johnston}, {Sheldon}, {Nichol},
  {Miller}, {Scranton}, {Bahcall}, {Barentine}, {Brewington}, {Brinkmann},
  {Harvanek}, {Kleinman}, {Krzesinski}, {Long}, {Nitta}, {Schneider},
  {Sneddin}, {Voges}, \& {York}}]{koester}
{Koester} B.~P. {et~al.}, 2007, \apj, 660, 239

\bibitem[{Koopmans, Owen \& Rosenblatt(1964)Koopmans, Owen, \&
  Rosenblatt}]{Koopmans1964}
Koopmans A. L.~H., Owen D.~B., Rosenblatt J.~I., 1964, Biometrika, 51, 25

\bibitem[{{Laigle} {et~al}\mbox{.}(2016){Laigle}, {McCracken}, {Ilbert},
  {Hsieh}, {Davidzon}, {Capak}, {Hasinger}, {Silverman}, {Pichon}, {Coupon},
  {Aussel}, {Le Borgne}, {Caputi}, {Cassata}, {Chang}, {Civano}, {Dunlop},
  {Fynbo}, {Kartaltepe}, {Koekemoer}, {Le F{\`e}vre}, {Le Floc'h}, {Leauthaud},
  {Lilly}, {Lin}, {Marchesi}, {Milvang-Jensen}, {Salvato}, {Sanders},
  {Scoville}, {Smolcic}, {Stockmann}, {Taniguchi}, {Tasca}, {Toft}, {Vaccari},
  \& {Zabl}}]{laigle}
{Laigle} C. {et~al.}, 2016, \apjs, 224, 24

\bibitem[{{Laureijs} {et~al}\mbox{.}(2011){Laureijs}, {Amiaux}, {Arduini},
  {Augu{\`e}res}, {Brinchmann}, {Cole}, {Cropper}, {Dabin}, {Duvet}, {Ealet},
  {Garilli}, {Gondoin}, {Guzzo}, {Hoar}, {Hoekstra}, {Holmes}, {Kitching},
  {Maciaszek}, {Mellier}, {Pasian}, {Percival}, {Rhodes}, {Saavedra Criado},
  {Sauvage}, {Scaramella}, {Valenziano}, {Warren}, {Bender}, {Castander},
  {Cimatti}, {Le F{\`e}vre}, {Kurki-Suonio}, {Levi}, {Lilje}, {Meylan},
  {Nichol}, {Pedersen}, {Popa}, {Rebolo Lopez}, {Rix}, {Rottgering},
  {Zeilinger}, {Grupp}, {Hudelot}, {Massey}, {Meneghetti}, {Miller}, {Paltani},
  {Paulin-Henriksson}, {Pires}, {Saxton}, {Schrabback}, {Seidel}, {Walsh},
  {Aghanim}, {Amendola}, {Bartlett}, {Baccigalupi}, {Beaulieu}, {Benabed},
  {Cuby}, {Elbaz}, {Fosalba}, {Gavazzi}, {Helmi}, {Hook}, {Irwin}, {Kneib},
  {Kunz}, {Mannucci}, {Moscardini}, {Tao}, {Teyssier}, {Weller}, {Zamorani},
  {Zapatero Osorio}, {Boulade}, {Foumond}, {Di Giorgio}, {Guttridge}, {James},
  {Kemp}, {Martignac}, {Spencer}, {Walton}, {Bl{\"u}mchen}, {Bonoli},
  {Bortoletto}, {Cerna}, {Corcione}, {Fabron}, {Jahnke}, {Ligori}, {Madrid},
  {Martin}, {Morgante}, {Pamplona}, {Prieto}, {Riva}, {Toledo}, {Trifoglio},
  {Zerbi}, {Abdalla}, {Douspis}, {Grenet}, {Borgani}, {Bouwens}, {Courbin},
  {Delouis}, {Dubath}, {Fontana}, {Frailis}, {Grazian}, {Koppenh{\"o}fer},
  {Mansutti}, {Melchior}, {Mignoli}, {Mohr}, {Neissner}, {Noddle}, {Poncet},
  {Scodeggio}, {Serrano}, {Shane}, {Starck}, {Surace}, {Taylor},
  {Verdoes-Kleijn}, {Vuerli}, {Williams}, {Zacchei}, {Altieri}, {Escudero
  Sanz}, {Kohley}, {Oosterbroek}, {Astier}, {Bacon}, {Bardelli}, {Baugh},
  {Bellagamba}, {Benoist}, {Bianchi}, {Biviano}, {Branchini}, {Carbone},
  {Cardone}, {Clements}, {Colombi}, {Conselice}, {Cresci}, {Deacon}, {Dunlop},
  {Fedeli}, {Fontanot}, {Franzetti}, {Giocoli}, {Garcia-Bellido}, {Gow},
  {Heavens}, {Hewett}, {Heymans}, {Holland}, {Huang}, {Ilbert}, {Joachimi},
  {Jennins}, {Kerins}, {Kiessling}, {Kirk}, {Kotak}, {Krause}, {Lahav}, {van
  Leeuwen}, {Lesgourgues}, {Lombardi}, {Magliocchetti}, {Maguire}, {Majerotto},
  {Maoli}, {Marulli}, {Maurogordato}, {McCracken}, {McLure}, {Melchiorri},
  {Merson}, {Moresco}, {Nonino}, {Norberg}, {Peacock}, {Pello}, {Penny},
  {Pettorino}, {Di Porto}, {Pozzetti}, {Quercellini}, {Radovich}, {Rassat},
  {Roche}, {Ronayette}, {Rossetti}, {Sartoris}, {Schneider}, {Semboloni},
  {Serjeant}, {Simpson}, {Skordis}, {Smadja}, {Smartt}, {Spano}, {Spiro},
  {Sullivan}, {Tilquin}, {Trotta}, {Verde}, {Wang}, {Williger}, {Zhao},
  {Zoubian}, \& {Zucca}}]{euclid}
{Laureijs} R. {et~al.}, 2011, arXiv e-prints, arXiv:1110.3193

\bibitem[{{Lieu} {et~al}\mbox{.}(2016){Lieu}, {Smith}, {Giles}, {Ziparo},
  {Maughan}, {D{\'e}mocl{\`e}s}, {Pacaud}, {Pierre}, {Adami}, {Bah{\'e}},
  {Clerc}, {Chiappetti}, {Eckert}, {Ettori}, {Lavoie}, {Le Fevre}, {McCarthy},
  {Kilbinger}, {Ponman}, {Sadibekova}, \& {Willis}}]{2016A&A...592A...4L}
{Lieu} M. {et~al.}, 2016, \aap, 592, A4

\bibitem[{{Lloyd-Davies} {et~al}\mbox{.}(2011){Lloyd-Davies}, {Romer},
  {Mehrtens}, {Hosmer}, {Davidson}, {Sabirli}, {Mann}, {Hilton}, {Liddle},
  {Viana}, {Campbell}, {Collins}, {Dubois}, {Freeman}, {Harrison}, {Hoyle},
  {Kay}, {Kuwertz}, {Miller}, {Nichol}, {Sahl{\'e}n}, {Stanford}, \&
  {Stott}}]{LD11}
{Lloyd-Davies} E.~J. {et~al.}, 2011, \mnras, 418, 14

\bibitem[{{Logan} {et~al}\mbox{.}(2018){Logan}, {Maughan}, {Bremer}, {Giles},
  {Birkinshaw}, {Chiappetti}, {Clerc}, {Faccioli}, {Koulouridis}, {Pacaud},
  {Pierre}, {Ramos-Ceja}, {Vignali}, \& {Willis}}]{Logan19}
{Logan} C.~H.~A. {et~al.}, 2018, \aap, 620, A18

\bibitem[{{Mahajan} \& {Raychaudhury}(2009)}]{2009MNRAS.400..687M}
{Mahajan} S., {Raychaudhury} S., 2009, \mnras, 400, 687

\bibitem[{{Mahdavi} {et~al}\mbox{.}(2013){Mahdavi}, {Hoekstra}, {Babul},
  {Bildfell}, {Jeltema}, \& {Henry}}]{2013ApJ...767..116M}
{Mahdavi} A., {Hoekstra} H., {Babul} A., {Bildfell} C., {Jeltema} T., {Henry}
  J.~P., 2013, \apj, 767, 116

\bibitem[{{Mantz} {et~al}\mbox{.}(2015){Mantz}, {Allen}, {Morris}, {Schmidt},
  {von der Linden}, \& {Urban}}]{Mantz:2015:wtgI}
{Mantz} A.~B., {Allen} S.~W., {Morris} R.~G., {Schmidt} R.~W., {von der Linden}
  A., {Urban} O., 2015, \mnras, 449, 199

\bibitem[{{Mantz} {et~al}\mbox{.}(2016){Mantz}, {Allen}, {Morris}, {von der
  Linden}, {Applegate}, {Kelly}, {Burke}, {Donovan}, \&
  {Ebeling}}]{Mantz:2016WtG-V}
{Mantz} A.~B. {et~al.}, 2016, \mnras, 463, 3582

\bibitem[{{Marigo} \& {Girardi}(2007)}]{padova2}
{Marigo} P., {Girardi} L., 2007, \aap, 469, 239

\bibitem[{{Marigo} {et~al}\mbox{.}(2008){Marigo}, {Girardi}, {Bressan},
  {Groenewegen}, {Silva}, \& {Granato}}]{padova3}
{Marigo} P., {Girardi} L., {Bressan} A., {Groenewegen} M.~A.~T., {Silva} L.,
  {Granato} G.~L., 2008, \aap, 482, 883

\bibitem[{{Melchior} {et~al}\mbox{.}(2016){Melchior}, {Gruen}, {McClintock},
  {Varga}, {Sheldon}, {Rozo}, {Amara}, {Becker}, {Benson}, {Bermeo}, {Bridle},
  {Clampitt}, {Dietrich}, {Hartley}, {Hollowood}, {Jain}, {Jarvis}, {Jeltema},
  {Kacprzak}, {MacCrann}, {Rykoff}, {Saro}, {Suchyta}, {Troxel}, {Zuntz},
  {Bonnett}, {Plazas}, {Abbott}, {Abdalla}, {Annis}, {Benoit-L{\'e}vy},
  {Bernstein}, {Bertin}, {Brooks}, {Buckley-Geer}, {Carnero Rosell}, {Carrasco
  Kind}, {Carretero}, {Cunha}, {D'Andrea}, {da Costa}, {Desai}, {Eifler},
  {Flaugher}, {Fosalba}, {Garc{\'{\i}}a-Bellido}, {Gaztanaga}, {Gerdes},
  {Gruendl}, {Gschwend}, {Gutierrez}, {Honscheid}, {James}, {Kirk}, {Krause},
  {Kuehn}, {Kuropatkin}, {Lahav}, {Lima}, {Maia}, {March}, {Martini},
  {Menanteau}, {Miller}, {Miquel}, {Mohr}, {Nichol}, {Ogando}, {Romer},
  {Sanchez}, {Scarpine}, {Sevilla-Noarbe}, {Smith}, {Soares-Santos},
  {Sobreira}, {Swanson}, {Tarle}, {Thomas}, {Walker}, {Weller}, \&
  {Zhang}}]{melchior}
{Melchior} P. {et~al.}, 2016, ArXiv e-prints

\bibitem[{{Miller} {et~al}\mbox{.}(2005){Miller}, {Nichol}, {Reichart},
  {Wechsler}, {Evrard}, {Annis}, {McKay}, {Bahcall}, {Bernardi}, {Boehringer},
  {Connolly}, {Goto}, {Kniazev}, {Lamb}, {Postman}, {Schneider}, {Sheth}, \&
  {Voges}}]{miller05}
{Miller} C.~J. {et~al.}, 2005, \aj, 130, 968

\bibitem[{{Mitchell} {et~al}\mbox{.}(2013){Mitchell}, {Lacey}, {Baugh}, \&
  {Cole}}]{mitchell}
{Mitchell} P.~D., {Lacey} C.~G., {Baugh} C.~M., {Cole} S., 2013, \mnras, 435,
  87

\bibitem[{{Mohr} {et~al}\mbox{.}(2012){Mohr}, {Armstrong}, {Bertin}, {Daues},
  {Desai}, {Gower}, {Gruendl}, {Hanlon}, {Kuropatkin}, {Lin}, {Marriner},
  {Petravic}, {Sevilla}, {Swanson}, {Tomashek}, {Tucker}, \&
  {Yanny}}]{dataproc}
{Mohr} J.~J. {et~al.}, 2012, Society of Photo-Optical Instrumentation Engineers
  (SPIE) Conference Series, 8451, 0

\bibitem[{{Mulroy} {et~al}\mbox{.}(2019){Mulroy}, {Farahi}, {Evrard}, {Smith},
  {Finoguenov}, {O'Donnell}, {Marrone}, {Abdulla}, {Bourdin}, {Carlstrom},
  {D{\'e}mocl{\`e}s}, {Haines}, {Martino}, {Mazzotta}, {McGee}, \&
  {Okabe}}]{mulroy2019}
{Mulroy} S.~L. {et~al.}, 2019, \mnras

\bibitem[{{Navarro}, {Frenk} \& {White}(1996){Navarro}, {Frenk}, \&
  {White}}]{nfw}
{Navarro} J.~F., {Frenk} C.~S., {White} S.~D.~M., 1996, \apj, 462, 563

\bibitem[{{Oemler}(1974)}]{oemler}
{Oemler}, Jr. A., 1974, \apj, 194, 1

\bibitem[{{Oguri}(2014)}]{oguri}
{Oguri} M., 2014, \mnras, 444, 147

\bibitem[{{Palmese} {et~al}\mbox{.}(2019){Palmese}, {Annis}, {Li}, {Welch},
  {Soares-Santos}, {Palmese}, {Annis}, {Li}, {Welch}, {Soares-Santos}, {Annis},
  {Li}, {Welch}, \& {Soares-Santos}}]{palmese19}
{Palmese} A. {et~al.}, 2019, in prep.

\bibitem[{{Palmese} {et~al}\mbox{.}(2016){Palmese}, {Lahav}, {Banerji},
  {Gruen}, {Jouvel}, {Melchior}, {Aleksi{\'c}}, {Annis}, {Diehl}, {Hartley},
  {Jeltema}, {Romer}, {Rozo}, {Rykoff}, {Seitz}, {Suchyta}, {Zhang}, {Abbott},
  {Abdalla}, {Allam}, {Benoit-L{\'e}vy}, {Bertin}, {Brooks}, {Buckley-Geer},
  {Burke}, {Capozzi}, {Carnero Rosell}, {Carrasco Kind}, {Carretero}, {Crocce},
  {Cunha}, {D'Andrea}, {da Costa}, {Desai}, {Dietrich}, {Doel}, {Estrada},
  {Evrard}, {Flaugher}, {Frieman}, {Gerdes}, {Goldstein}, {Gruendl},
  {Gutierrez}, {Honscheid}, {James}, {Kuehn}, {Kuropatkin}, {Li}, {Lima},
  {Maia}, {Marshall}, {Miller}, {Miquel}, {Nord}, {Ogando}, {Plazas},
  {Roodman}, {Sanchez}, {Scarpine}, {Sevilla-Noarbe}, {Smith}, {Soares-Santos},
  {Sobreira}, {Swanson}, {Tarle}, {Thomas}, {Tucker}, \& {Vikram}}]{palmese}
{Palmese} A. {et~al.}, 2016, \mnras, 463, 1486

\bibitem[{{Pereira} {et~al}\mbox{.}(2018){Pereira}, {Soares-Santos}, {Makler},
  {Annis}, {Lin}, {Palmese}, {Vitorelli}, {Welch}, {Caminha}, {Erben},
  {Moraes}, \& {Shan}}]{maria}
{Pereira} M.~E.~S. {et~al.}, 2018, \mnras, 474, 1361

\bibitem[{{Press} \& {Schechter}(1974)}]{press}
{Press} W.~H., {Schechter} P., 1974, \apj, 187, 425

\bibitem[{{Rozo} {et~al}\mbox{.}(2011){Rozo}, {Rykoff}, {Koester}, {Nord},
  {Wu}, {Evrard}, \& {Wechsler}}]{extrinsicscatter}
{Rozo} E., {Rykoff} E., {Koester} B., {Nord} B., {Wu} H.-Y., {Evrard} A.,
  {Wechsler} R., 2011, \apj, 740, 53

\bibitem[{{Rozo} {et~al}\mbox{.}(2009{\natexlab{a}}){Rozo}, {Rykoff}, {Evrard},
  {Becker}, {McKay}, {Wechsler}, {Koester}, {Hao}, {Hansen}, {Sheldon},
  {Johnston}, {Annis}, \& {Frieman}}]{rozo09}
{Rozo} E. {et~al.}, 2009{\natexlab{a}}, \apj, 699, 768

\bibitem[{{Rozo} {et~al}\mbox{.}(2009{\natexlab{b}}){Rozo}, {Rykoff},
  {Koester}, {McKay}, {Hao}, {Evrard}, {Wechsler}, {Hansen}, {Sheldon},
  {Johnston}, {Becker}, {Annis}, {Bleem}, \& {Scranton}}]{lambda}
{Rozo} E. {et~al.}, 2009{\natexlab{b}}, \apj, 703, 601

\bibitem[{{Rykoff} {et~al}\mbox{.}(2014){Rykoff}, {Rozo}, {Busha}, {Cunha},
  {Finoguenov}, {Evrard}, {Hao}, {Koester}, {Leauthaud}, {Nord}, {Pierre},
  {Reddick}, {Sadibekova}, {Sheldon}, \& {Wechsler}}]{redpaper}
{Rykoff} E.~S. {et~al.}, 2014, \apj, 785, 104

\bibitem[{{Rykoff} {et~al}\mbox{.}(2016){Rykoff}, {Rozo}, {Hollowood},
  {Bermeo-Hernandez}, {Jeltema}, {Mayers}, {Romer}, {Rooney}, {Saro}, {Vergara
  Cervantes}, {Wechsler}, {Wilcox}, {Abbott}, {Abdalla}, {Allam}, {Annis},
  {Benoit-L{\'e}vy}, {Bernstein}, {Bertin}, {Brooks}, {Burke}, {Capozzi},
  {Carnero Rosell}, {Carrasco Kind}, {Castander}, {Childress}, {Collins},
  {Cunha}, {D'Andrea}, {da Costa}, {Davis}, {Desai}, {Diehl}, {Dietrich},
  {Doel}, {Evrard}, {Finley}, {Flaugher}, {Fosalba}, {Frieman}, {Glazebrook},
  {Goldstein}, {Gruen}, {Gruendl}, {Gutierrez}, {Hilton}, {Honscheid}, {Hoyle},
  {James}, {Kay}, {Kuehn}, {Kuropatkin}, {Lahav}, {Lewis}, {Lidman}, {Lima},
  {Maia}, {Mann}, {Marshall}, {Martini}, {Melchior}, {Miller}, {Miquel},
  {Mohr}, {Nichol}, {Nord}, {Ogando}, {Plazas}, {Reil}, {Sahl{\'e}n},
  {Sanchez}, {Santiago}, {Scarpine}, {Schubnell}, {Sevilla-Noarbe}, {Smith},
  {Soares-Santos}, {Sobreira}, {Stott}, {Suchyta}, {Swanson}, {Tarle},
  {Thomas}, {Tucker}, {Uddin}, {Viana}, {Vikram}, {Walker}, {Zhang}, \& {DES
  Collaboration}}]{redmappersv}
{Rykoff} E.~S. {et~al.}, 2016, \apjs, 224, 1

\bibitem[{{S{\'a}nchez-Bl{\'a}zquez}
  {et~al}\mbox{.}(2006){S{\'a}nchez-Bl{\'a}zquez}, {Peletier},
  {Jim{\'e}nez-Vicente}, {Cardiel}, {Cenarro}, {Falc{\'o}n-Barroso}, {Gorgas},
  {Selam}, \& {Vazdekis}}]{miles}
{S{\'a}nchez-Bl{\'a}zquez} P. {et~al.}, 2006, \mnras, 371, 703

\bibitem[{{Sevilla} {et~al}\mbox{.}(2011){Sevilla}, {Armstrong}, {Bertin},
  {Carlson}, {Daues}, {Desai}, {Gower}, {Gruendl}, {Hanlon}, {Jarvis},
  {Kessler}, {Kuropatkin}, {Lin}, {Marriner}, {Mohr}, {Petravick}, {Sheldon},
  {Swanson}, {Tomashek}, {Tucker}, {Yang}, {Yanny}, \& {for the DES
  Collaboration}}]{sevilla}
{Sevilla} I. {et~al.}, 2011, preprint (arXiv:astro-ph/1109.6741)

\bibitem[{{Sheth} \& {Tormen}(2002)}]{sheth}
{Sheth} R.~K., {Tormen} G., 2002, \mnras, 329, 61

\bibitem[{{Simet} {et~al}\mbox{.}(2017){Simet}, {McClintock}, {Mandelbaum},
  {Rozo}, {Rykoff}, {Sheldon}, \& {Wechsler}}]{simet}
{Simet} M., {McClintock} T., {Mandelbaum} R., {Rozo} E., {Rykoff} E., {Sheldon}
  E., {Wechsler} R.~H., 2017, \mnras, 466, 3103

\bibitem[{{Simha} {et~al}\mbox{.}(2014){Simha}, {Weinberg}, {Conroy}, {Dave},
  {Fardal}, {Katz}, \& {Oppenheimer}}]{simha}
{Simha} V., {Weinberg} D.~H., {Conroy} C., {Dave} R., {Fardal} M., {Katz} N.,
  {Oppenheimer} B.~D., 2014, ArXiv e-prints

\bibitem[{{Soares-Santos} {et~al}\mbox{.}(2011){Soares-Santos}, {de Carvalho},
  {Annis}, {Gal}, {La Barbera}, {Lopes}, {Wechsler}, {Busha}, \& {Gerke}}]{vt}
{Soares-Santos} M. {et~al.}, 2011, \apj, 727, 45

\bibitem[{{Taylor} {et~al}\mbox{.}(2011){Taylor}, {Hopkins}, {Baldry}, {Brown},
  {Driver}, {Kelvin}, {Hill}, {Robotham}, {Bland-Hawthorn}, {Jones}, {Sharp},
  {Thomas}, {Liske}, {Loveday}, {Norberg}, {Peacock}, {Bamford}, {Brough},
  {Colless}, {Cameron}, {Conselice}, {Croom}, {Frenk}, {Gunawardhana},
  {Kuijken}, {Nichol}, {Parkinson}, {Phillipps}, {Pimbblet}, {Popescu},
  {Prescott}, {Sutherland}, {Tuffs}, {van Kampen}, \& {Wijesinghe}}]{taylor}
{Taylor} E.~N. {et~al.}, 2011, \mnras, 418, 1587

\bibitem[{{Taylor}(2005)}]{topcat}
{Taylor} M.~B., 2005, in Astronomical Society of the Pacific Conference Series,
  Vol. 347, Astronomical Data Analysis Software and Systems XIV, {Shopbell} P.,
  {Britton} M., {Ebert} R., eds., p.~29

\bibitem[{{The Dark Energy Survey Collaboration}(2005)}]{descollaboration}
{The Dark Energy Survey Collaboration}, 2005, preprint (arXiv:astro-ph/0510346)

\bibitem[{{Tinker} {et~al}\mbox{.}(2008){Tinker}, {Kravtsov}, {Klypin},
  {Abazajian}, {Warren}, {Yepes}, {Gottl{\"o}ber}, \& {Holz}}]{tinker}
{Tinker} J., {Kravtsov} A.~V., {Klypin} A., {Abazajian} K., {Warren} M.,
  {Yepes} G., {Gottl{\"o}ber} S., {Holz} D.~E., 2008, \apj, 688, 709

\bibitem[{{Wechsler} \& {Tinker}(2018)}]{2018arXiv180403097W}
{Wechsler} R.~H., {Tinker} J.~L., 2018, ARA\&A, 56, 435

\bibitem[{{Welch} {et~al}\mbox{.}(2019){Welch}, {Annis}, {Li}, {Palmese}, \&
  {Soares-Santos}}]{welch}
{Welch} B., {Annis} J., {Li} H., {Palmese} A., {Soares-Santos} M., 2019, in
  prep.

\bibitem[{{Zhang} {et~al}\mbox{.}(2017){Zhang}, {Miller}, {Rooney}, {Bermeo},
  {Romer}, {Vergara cervantes}, {Rykoff}, {Hennig}, {Das}, {Mckay}, {Song},
  {Wilcox}, {Bacon}, {Bridle}, {Collins}, {Conselice}, {Hilton}, {Hoyle},
  {Kay}, {Liddle}, {Mann}, {Mehrtens}, {Mayers}, {Nichol}, {Sahlen}, {Stott},
  {Viana}, {Wechsler}, {Abbott}, {Abdalla}, {Allam}, {Benoit-levy}, {Brooks},
  {Buckley-geer}, {Burke}, {Carnero rosell}, {Carrasco kind}, {Carretero},
  {Castander}, {Crocce}, {Cunha}, {Dandrea}, {Da costa}, {Diehl}, {Dietrich},
  {Eifler}, {Flaugher}, {Fosalba}, {Garcia-bellido}, {Gaztanaga}, {Gerdes},
  {Gruen}, {Gruendl}, {Gschwend}, {Gutierrez}, {Honscheid}, {James}, {Jeltema},
  {Kuehn}, {Kuropatkin}, {Lima}, {Lin}, {Maia}, {March}, {Marshall},
  {Melchior}, {Menanteau}, {Miquel}, {Ogando}, {Plazas}, {Sanchez},
  {Schubnell}, {Sevilla-noarbe}, {Smith}, {Soares-santos}, {Sobreira},
  {Suchyta}, {Swanson}, {Tarle}, \& {Walker}}]{zhangII}
{Zhang} Y. {et~al.}, 2017, arXiv e-prints

\bibitem[{{Zu} \& {Mandelbaum}(2016)}]{zu2}
{Zu} Y., {Mandelbaum} R., 2016, \mnras, 457, 4360

\end{thebibliography}

\newpage
\appendix
\section{Affiliations}
$^{1}$ Fermi National Accelerator Laboratory, P. O. Box 500, Batavia, IL 60510, USA \\$^{2}$ Department of Physics \& Astronomy, University College London, Gower Street, London, WC1E 6BT, UK \\$^{3}$ Department of Astronomy, The Ohio State University, Columbus, OH 43210, USA \\$^{4}$ Department of Physics, Carnegie Mellon University, Pittsburgh, Pennsylvania 15312, USA \\$^{5}$ Brandeis University, Physics Department, 415 South Street, Waltham MA 02453 \\$^{6}$ Kavli Institute for Cosmological Physics, University of Chicago, Chicago, IL 60637, USA \\$^{7}$ Department of Physics and Astronomy, Pevensey Building, University of Sussex, Brighton, BN1 9QH, UK \\$^{8}$ Santa Cruz Institute for Particle Physics, Santa Cruz, CA 95064, USA \\$^{9}$ Department of Astronomy, University of Michigan, Ann Arbor, MI 48109, USA \\$^{10}$ Department of Physics, University of Michigan, Ann Arbor, MI 48109, USA \\$^{11}$ Astrophysics \& Cosmology Research Unit, School of Mathematics, Statistics \& Computer Science, University of KwaZulu-Natal, Westville Campus, Durban 4041, South Africa \\$^{12}$ Department of Physics, Stanford University, 382 Via Pueblo Mall, Stanford, CA 94305, USA \\$^{13}$ Kavli Institute for Particle Astrophysics \& Cosmology, P. O. Box 2450, Stanford University, Stanford, CA 94305, USA \\$^{14}$ SLAC National Accelerator Laboratory, Menlo Park, CA 94025, USA \\$^{15}$ Department of Physics, ETH Zurich, Wolfgang-Pauli-Strasse 16, CH-8093 Zurich, Switzerland \\$^{16}$ New York University, CCPP,  New York, NY 10003, USA \\$^{17}$ Department of Physics, University of Arizona, Tucson, AZ 85721, USA \\$^{18}$ Max Planck Institute for Extraterrestrial Physics, Giessenbachstrasse, 85748 Garching, Germany \\$^{19}$ Universit\"ats-Sternwarte, Fakult\"at f\"ur Physik, Ludwig-Maximilians Universit\"at M\"unchen, Scheinerstr. 1, 81679 M\"unchen, Germany \\$^{20}$ Institute of Cosmology and Gravitation, University of Portsmouth, Portsmouth, PO1 3FX, UK \\$^{21}$ Centro de Investigaciones Energ\'eticas, Medioambientales y Tecnol\'ogicas (CIEMAT), Madrid, Spain \\$^{22}$ Laborat\'orio Interinstitucional de e-Astronomia - LIneA, Rua Gal. Jos\'e Cristino 77, Rio de Janeiro, RJ - 20921-400, Brazil \\$^{23}$ Department of Astronomy, University of Illinois at Urbana-Champaign, 1002 W. Green Street, Urbana, IL 61801, USA \\$^{24}$ National Center for Supercomputing Applications, 1205 West Clark St., Urbana, IL 61801, USA \\$^{25}$ Institut de F\'{\i}sica d'Altes Energies (IFAE), The Barcelona Institute of Science and Technology, Campus UAB, 08193 Bellaterra (Barcelona) Spain \\$^{26}$ Institut d'Estudis Espacials de Catalunya (IEEC), 08034 Barcelona, Spain \\$^{27}$ Institute of Space Sciences (ICE, CSIC),  Campus UAB, Carrer de Can Magrans, s/n,  08193 Barcelona, Spain \\$^{28}$ Astrophysics Research Institute, Liverpool John Moores University, IC2, Liverpool Science Park, 146 Brownlow Hill, Liverpool, L3 5RF, UK \\$^{29}$ Observat\'orio Nacional, Rua Gal. Jos\'e Cristino 77, Rio de Janeiro, RJ - 20921-400, Brazil \\$^{30}$ Department of Physics, IIT Hyderabad, Kandi, Telangana 502285, India \\$^{31}$ Excellence Cluster Origins, Boltzmannstr.\ 2, 85748 Garching, Germany \\$^{32}$ Faculty of Physics, Ludwig-Maximilians-Universit\"at, Scheinerstr. 1, 81679 Munich, Germany \\$^{33}$ Instituto de Fisica Teorica UAM/CSIC, Universidad Autonoma de Madrid, 28049 Madrid, Spain \\$^{34}$ Center for Cosmology and Astro-Particle Physics, The Ohio State University, Columbus, OH 43210, USA \\$^{35}$ Department of Physics, The Ohio State University, Columbus, OH 43210, USA \\$^{36}$ Harvard-Smithsonian Center for Astrophysics, Cambridge, MA 02138, USA \\$^{37}$ Department of Astronomy/Steward Observatory, University of Arizona, 933 North Cherry Avenue, Tucson, AZ 85721-0065, USA \\$^{38}$ Australian Astronomical Optics, Macquarie University, North Ryde, NSW 2113, Australia \\$^{39}$ Institute for Astronomy, University of Edinburgh, Royal Observatory, Blackford Hill, Edinburgh, EH9 3HJ, UK \\$^{40}$ Departamento de F\'isica Matem\'atica, Instituto de F\'isica, Universidade de S\~ao Paulo, CP 66318, S\~ao Paulo, SP, 05314-970, Brazil \\$^{41}$ Institute for Astronomy, University of Edinburgh, Royal Observatory, Blackford Hill, Edinburgh EH9 3NJ \\$^{42}$ George P. and Cynthia Woods Mitchell Institute for Fundamental Physics and Astronomy, and Department of Physics and Astronomy, Texas A\&M University, College Station, TX 77843,  USA \\$^{43}$ Instituci\'o Catalana de Recerca i Estudis Avan\c{c}ats, E-08010 Barcelona, Spain \\$^{44}$ Department of Astrophysical Sciences, Princeton University, Peyton Hall, Princeton, NJ 08544, USA \\$^{45}$ BIPAC, Department of Physics, University of Oxford, Denys Wilkinson Building, 1 Keble Road, Oxford OX1 3RH, UK \\$^{46}$ Instituto de F\'isica Gleb Wataghin, Universidade Estadual de Campinas, 13083-859, Campinas, SP, Brazil \\$^{47}$ Sub-department of Astrophysics, Department of Physics, University of Oxford, Denys Wilkinson Building, Keble Road, Oxford OX1 3RH, UK and Department of Physics, Lancaster University, Lancaster LA1 4 YB, UK \\$^{48}$ Computer Science and Mathematics Division, Oak Ridge National Laboratory, Oak Ridge, TN 37831 \\$^{49}$ Departamento de F\'isica e Astronomia, Faculdade de Ciencias, Universidade do Porto, Rua do Campo Alegre, 687, P-4169-007 Porto, Portugal \\$^{50}$ Instituto de Astrof\'isica e Ciencias do Espaco, Universidade do Porto, CAUP, Rua das Estrelas, P-4150-762 Porto, Portugal \\$^{51}$ Argonne National Laboratory, 9700 South Cass Avenue, Lemont, IL 60439, USA \\$^{52}$ Cerro Tololo Inter-American Observatory, National Optical Astronomy Observatory, Casilla 603, La Serena, Chile \\

\newpage

\section{Cluster Catalogues} \label{app:cluster-cat}

In Table \ref{tab:chandra} and \ref{tab:XMM-table}, we provide the optical and X-ray properties of \emph{Chandra} and \emph{XMM} clusters. $z_{\lambda}$ is the cluster photometric redshift computed by redMaPPer. $\mu_\star$ is the mass proxy computed in this work, while the redMaPPer richness is $\lambda$.  The full redMaPPer DES Y1A1 catalogues will be available at \url{http://risa.stanford.edu/redmapper/}. \texttt{XCS\_NAME} in Table \ref{tab:XMM-table} is the unique source identifier which could be used to match with the XCS source catalog (Giles et al., in prep.).

\begin{table*}
\caption{\emph{Chandra} Clusters.} \label{tab:chandra}
\begin{tabular}{|l|l|l|l|l|}
\hline
$z_{\lambda}$ & $\lambda$ & $\mu_\star$ & ${\rm k}T_X \ [{\rm keV}]$ & obsid(s) \\ 
\hline
0.304 & $ 200.65 \pm  6.90$ & $2265.61 \pm 193.89$ & $ 10.90^{+  0.84}_{-  0.81}$ & 9331,15099 \\
 0.419 & $ 171.91 \pm  4.49$ & $2393.20 \pm 181.40$ & $  7.39^{+  0.41}_{-  0.32}$ & 13396,16355,17536 \\ 
 0.301 & $ 144.10 \pm  4.00$ & $2133.15 \pm 192.72$ & $ 10.24^{+  0.26}_{-  0.26}$ & 12260,16127,16282,16524,16525,16526 \\ 
 0.351 & $ 188.40 \pm 10.06$ & $2465.34 \pm 212.46$ & $ 14.89^{+  0.59}_{-  0.55}$ & 4966 \\ 
 0.368 & $ 138.85 \pm  4.72$ & $1381.93 \pm 148.01$ & $  7.88^{+  1.08}_{-  0.80}$ & 13395 \\ 
 0.240 & $ 136.44 \pm  4.69$ & $2207.50 \pm 235.86$ & $ 12.16^{+  1.36}_{-  0.92}$ & 15097 \\ 
 0.326 & $ 142.26 \pm  6.28$ & $2066.69 \pm 217.81$ & $  9.48^{+  0.73}_{-  0.53}$ & 11710,16285 \\ 
 0.526 & $ 159.39 \pm  6.29$ & $1906.99 \pm 120.57$ & $  7.51^{+  3.04}_{-  1.41}$ & 13466 \\ 
 0.278 & $ 132.86 \pm  4.36$ & $1743.45 \pm 193.26$ & $  9.46^{+  0.66}_{-  0.45}$ & 3248,11728 \\ 
 0.605 & $ 165.92 \pm  5.63$ & $1580.90 \pm 130.39$ & $  7.71^{+  0.84}_{-  0.55}$ & 12264,13116,13117 \\ 
 0.282 & $ 130.37 \pm  4.73$ & $941.93 \pm 122.49$ & $  6.25^{+  2.57}_{-  1.30}$ & 17162,16271,17162 \\ 
 0.591 & $ 152.04 \pm  5.00$ & $2074.08 \pm 194.61$ & $ 14.32^{+  0.52}_{-  0.52}$ & 13401,16135,16545 \\ 
 0.418 & $ 124.99 \pm  4.26$ & $1522.08 \pm 296.48$ & $ 11.83^{+  1.25}_{-  0.90}$ & 12259 \\ 
 0.231 & $ 135.36 \pm  6.69$ & $1769.49 \pm 166.19$ & $  9.88^{+  0.79}_{-  0.66}$ & 15108 \\ 
 0.348 & $ 125.67 \pm  5.83$ & $1186.05 \pm 136.16$ & $  5.86^{+  0.63}_{-  0.37}$ & 17185 \\ 
 0.425 & $ 130.46 \pm  6.13$ & $1601.02 \pm 119.29$ & $  7.00^{+  1.13}_{-  0.77}$ & 13463 \\ 
 0.443 & $ 138.20 \pm  6.42$ & $1216.62 \pm 118.77$ & $  8.67^{+  1.20}_{-  1.07}$ & 13402 \\ 
 0.107 & $ 98.49 \pm  4.57$ & $1556.41 \pm 153.54$ & $  6.90^{+  0.33}_{-  0.33}$ & 15313 \\ 
 0.343 & $ 121.80 \pm  5.16$ & $1713.03 \pm 162.38$ & $  8.93^{+  2.15}_{-  1.36}$ & 12269 \\ 
 0.637 & $ 137.04 \pm  5.68$ & $1629.48 \pm 175.04$ & $  6.26^{+  0.86}_{-  0.61}$ & 13491 \\ 
 0.485 & $ 112.56 \pm  4.59$ & $1425.71 \pm 218.74$ & $  9.81^{+  1.63}_{-  1.05}$ & 16230 \\ 
 0.482 & $ 124.29 \pm  5.23$ & $1851.69 \pm 128.73$ & $  9.02^{+  1.65}_{-  1.18}$ & 13398 \\ 
 0.536 & $ 118.49 \pm  4.75$ & $1362.55 \pm 116.42$ & $ 10.25^{+  6.73}_{-  2.71}$ & 9416 \\ 
 0.405 & $ 99.65 \pm  3.90$ & $1175.95 \pm 102.05$ & $  6.79^{+  0.61}_{-  0.57}$ & 12265 \\ 
 0.585 & $ 103.10 \pm  4.53$ & $1319.48 \pm 216.97$ & $  7.03^{+  0.60}_{-  0.55}$ & 13397 \\ 
 0.401 & $ 99.89 \pm  3.76$ & $1163.74 \pm  93.37$ & $  8.33^{+  1.09}_{-  0.96}$ & 13403 \\ 
 0.282 & $ 94.42 \pm  3.58$ & $1045.47 \pm 131.45$ & $  7.58^{+  0.79}_{-  0.59}$ & 12262 \\ 
 0.261 & $ 97.13 \pm  5.50$ & $872.69 \pm 130.77$ & $  6.29^{+  0.28}_{-  0.28}$ & 5786,17170,17490,18702,18703 \\ 
 0.357 & $ 89.86 \pm  3.96$ & $1073.72 \pm 132.79$ & $  6.18^{+  0.98}_{-  0.80}$ & 13465 \\ 
 0.531 & $ 107.99 \pm  5.15$ & $791.36 \pm  79.59$ & $  4.84^{+  0.69}_{-  0.53}$ & 12270,13155 \\ 
 0.337 & $ 82.20 \pm  3.45$ & $806.08 \pm 131.55$ & $  5.06^{+  0.51}_{-  0.49}$ & 12266 \\ 
 0.317 & $ 93.97 \pm  4.67$ & $948.72 \pm 103.81$ & $  4.70^{+  0.50}_{-  0.37}$ & 11998 \\ 
 0.290 & $ 133.10 \pm 10.07$ & $1055.20 \pm  81.84$ & $ 11.31^{+  1.35}_{-  1.07}$ & 4993 \\ 
 0.207 & $ 84.93 \pm  4.41$ & $826.01 \pm 114.00$ & $  7.91^{+  0.55}_{-  0.50}$ & 15111 \\ 
 0.222 & $ 101.77 \pm  6.80$ & $867.17 \pm 111.58$ & $  6.79^{+  0.29}_{-  0.29}$ & 15110,17476 \\ 
 0.430 & $ 110.63 \pm  6.45$ & $880.08 \pm  83.50$ & $  5.60^{+  0.82}_{-  0.59}$ & 893 \\ 
 0.428 & $ 93.52 \pm  4.60$ & $1261.95 \pm 150.78$ & $  6.76^{+  0.92}_{-  0.81}$ & 13504 \\ 
 0.115 & $ 81.27 \pm  3.48$ & $1271.34 \pm 209.00$ & $ 10.29^{+  0.71}_{-  0.71}$ & 15304 \\ 
 0.188 & $ 72.33 \pm  3.17$ & $569.79 \pm  70.29$ & $  7.67^{+  0.54}_{-  0.45}$ & 15122 \\ 
 0.590 & $ 82.09 \pm  3.98$ & $803.66 \pm  99.32$ & $  7.11^{+  1.11}_{-  0.76}$ & 13481,14448 \\ 
 0.564 & $ 84.61 \pm  3.93$ & $1009.03 \pm 165.45$ & $  6.14^{+  1.91}_{-  1.17}$ & 12263 \\ 
 0.498 & $ 81.69 \pm  3.68$ & $1202.31 \pm 114.41$ & $ 11.56^{+  3.81}_{-  2.01}$ & 13487 \\ 
 0.649 & $ 113.97 \pm  6.46$ & $1248.34 \pm 125.15$ & $  7.82^{+  1.17}_{-  0.80}$ & 914 \\ 
 0.301 & $ 90.48 \pm  4.76$ & $1746.10 \pm 143.65$ & $  4.72^{+  0.58}_{-  0.37}$ & 4208 \\ 
 0.645 & $ 90.19 \pm  5.08$ & $1108.13 \pm 140.96$ & $  8.96^{+  1.18}_{-  0.85}$ & 13484,15573 \\ 
 0.313 & $ 73.50 \pm  3.26$ & $926.14 \pm 107.00$ & $  2.14^{+  0.48}_{-  0.33}$ & 11746 \\ 
 0.109 & $ 70.02 \pm  3.39$ & $1243.21 \pm 134.28$ & $  5.45^{+  0.36}_{-  0.34}$ & 4980 \\ 
 0.410 & $ 104.42 \pm  7.85$ & $724.34 \pm  88.97$ & $  7.13^{+  0.87}_{-  0.67}$ & 13462 \\ 
 0.246 & $ 75.56 \pm  5.47$ & $912.49 \pm 165.76$ & $  7.25^{+  0.18}_{-  0.18}$ & 552,9370 \\ 
 0.565 & $ 75.10 \pm  4.04$ & $935.19 \pm 189.65$ & $  8.16^{+  1.06}_{-  0.84}$ & 13476 \\ 
 0.153 & $ 61.86 \pm  3.46$ & $526.95 \pm  64.93$ & $  4.53^{+  0.16}_{-  0.16}$ & 9417,16280 \\ 
 0.629 & $ 73.29 \pm  4.06$ & $1189.79 \pm 242.52$ & $  8.51^{+  3.09}_{-  1.67}$ & 13472 \\ 
 0.568 & $ 88.73 \pm  5.90$ & $1095.82 \pm 158.79$ & $  8.11^{+  1.15}_{-  0.86}$ & 13474 \\ 
 0.471 & $ 69.12 \pm  3.47$ & $743.56 \pm  88.51$ & $  6.48^{+  0.72}_{-  0.62}$ & 13470 \\ 
 0.103 & $ 67.05 \pm  4.54$ & $547.98 \pm  73.66$ & $  6.14^{+  0.23}_{-  0.21}$ & 15107,15312 \\ 
 0.455 & $ 81.02 \pm  6.04$ & $751.97 \pm  98.22$ & $  9.85^{+  1.67}_{-  1.13}$ & 9432 \\ 
 0.114 & $ 64.35 \pm  3.38$ & $1121.34 \pm 128.50$ & $  4.90^{+  0.53}_{-  0.44}$ & 9383 \\ 
 0.657 & $ 65.40 \pm  3.67$ & $821.83 \pm 157.98$ & $  6.63^{+  1.36}_{-  0.93}$ & 13490,15571 \\ 
 0.414 & $ 60.15 \pm  3.46$ & $782.76 \pm 148.14$ & $  7.77^{+  1.22}_{-  1.00}$ & 13494 \\ 
 \end{tabular}
\end{table*}

\begin{table*}\caption*{continued.}
    \centering
\begin{tabular}{|l|l|l|l|l|l|}
 0.474 & $ 62.76 \pm  6.45$ & $465.57 \pm  67.00$ & $  5.64^{+  1.52}_{-  1.05}$ & 12268 \\ 
 0.310 & $ 34.28 \pm  2.80$ & $2848.21 \pm 245.38$ & $  7.36^{+  6.01}_{-  2.14}$ & 15107,15312 \\ 
 0.227 & $ 30.00 \pm  2.66$ & $247.16 \pm  82.65$ & $  3.41^{+  0.99}_{-  0.56}$ & 13484,15573 \\
 0.164 & $ 43.28 \pm  4.13$ & $584.58 \pm  84.78$ & $  6.30^{+  0.30}_{-  0.29}$ & 9416 \\ 
 0.403 & $ 23.65 \pm  2.21$ & $154.81 \pm  64.28$ & $  1.62^{+  0.62}_{-  0.27}$ & 16981 \\ 

\hline
\end{tabular}
\end{table*}

\begin{table*}
\caption{\emph{XMM} Clusters.} \label{tab:XMM-table}
\begin{tabular}{|l|l|l|l|l|l|}
\hline
\texttt{MEM\_MATCH\_ID} & $z_{\lambda}$ & $\lambda$ & $\mu_\star$ & ${\rm k}T_X \ [{\rm keV}]$ & XCS\_NAME \\ \hline
0.429 & $ 234.50 \pm  7.52$ & $2965.79 \pm 204.70$ & $  7.41^{+  7.59}_{-  7.24}$ & XMMXCS J025417.8-585705.2 \\ 
0.303 & $ 195.07 \pm  6.78$ & $2260.15 \pm 192.60$ & $  5.76^{+  5.86}_{-  5.66}$ & XMMXCS J051636.6-543120.8 \\ 
0.352 & $ 178.84 \pm  8.71$ & $2464.07 \pm 212.98$ & $ 10.28^{+ 10.44}_{- 10.12}$ & XMMXCS J224844.9-443141.7 \\ 
0.421 & $ 174.46 \pm  5.07$ & $2392.56 \pm 181.57$ & $  5.81^{+  6.87}_{-  4.98}$ & XMMXCS J041114.1-481910.9 \\ 
0.604 & $ 169.08 \pm  5.77$ & $1580.02 \pm 130.30$ & $  6.89^{+  7.60}_{-  6.28}$ & XMMXCS J055943.5-524937.5 \\ 
0.301 & $ 146.24 \pm  4.04$ & $2151.24 \pm 193.82$ & $  8.31^{+  8.64}_{-  8.00}$ & XMMXCS J024529.3-530210.7 \\ 
0.326 & $ 141.08 \pm  5.96$ & $2063.03 \pm 217.59$ & $  6.77^{+  7.67}_{-  6.04}$ & XMMXCS J213511.8-010258.0 \\ 
0.443 & $ 138.53 \pm  6.45$ & $1216.62 \pm 118.77$ & $  5.86^{+  6.10}_{-  5.64}$ & XMMXCS J030415.7-440153.0 \\ 
0.231 & $ 136.78 \pm  7.18$ & $1643.18 \pm 160.40$ & $  7.93^{+  8.50}_{-  7.41}$ & XMMXCS J202323.2-553504.7 \\ 
0.239 & $ 135.48 \pm  5.08$ & $2192.19 \pm 234.86$ & $  8.51^{+  8.78}_{-  8.26}$ & XMMXCS J213516.8+012600.0 \\ 
0.425 & $ 130.39 \pm  6.17$ & $1682.62 \pm 142.32$ & $  6.38^{+  7.94}_{-  5.25}$ & XMMXCS J213538.5-572616.6 \\ 
0.278 & $ 129.00 \pm  4.30$ & $1740.84 \pm 193.03$ & $  7.15^{+  7.37}_{-  6.94}$ & XMMXCS J233738.6+001614.5 \\ 
0.494 & $ 126.99 \pm  4.31$ & $1663.42 \pm 173.41$ & $  5.78^{+  5.98}_{-  5.59}$ & XMMXCS J024339.4-483338.3 \\ 
0.481 & $ 124.41 \pm  5.23$ & $1851.19 \pm 128.87$ & $  7.18^{+  7.77}_{-  6.66}$ & XMMXCS J214551.9-564453.6 \\ 
0.346 & $ 119.25 \pm  4.42$ & $1727.99 \pm 165.14$ & $  4.94^{+  5.36}_{-  4.58}$ & XMMXCS J021711.6-524512.9 \\ 
0.649 & $ 115.18 \pm  6.49$ & $1248.41 \pm 125.17$ & $  6.91^{+  7.83}_{-  6.11}$ & XMMXCS J054250.3-410003.5 \\ 
0.429 & $ 107.32 \pm  6.11$ & $878.31 \pm  83.43$ & $  4.33^{+  4.44}_{-  4.23}$ & XMMXCS J033044.8-522921.9 \\ 
0.584 & $ 104.70 \pm  4.61$ & $1318.88 \pm 216.92$ & $  7.03^{+  7.35}_{-  6.73}$ & XMMXCS J041722.7-474847.1 \\ 
0.410 & $ 103.78 \pm  7.82$ & $723.43 \pm  88.67$ & $  6.76^{+  8.47}_{-  5.60}$ & XMMXCS J001318.8-490651.9 \\ 
0.287 & $ 101.27 \pm  6.38$ & $1052.83 \pm  81.40$ & $  6.25^{+  6.39}_{-  6.11}$ & XMMXCS J023217.6-442053.8 \\ 
0.401 & $ 100.89 \pm  4.00$ & $1163.61 \pm  93.36$ & $  5.53^{+  5.73}_{-  5.34}$ & XMMXCS J023442.5-583121.0 \\ 
0.106 & $ 98.55 \pm  4.63$ & $1548.93 \pm 152.71$ & $  4.83^{+  4.92}_{-  4.74}$ & XMMXCS J032835.9-554239.3 \\ 
0.262 & $ 93.40 \pm  6.24$ & $881.17 \pm 131.71$ & $  5.45^{+  5.65}_{-  5.27}$ & XMMXCS J234341.7+001831.2 \\ 
0.523 & $ 92.45 \pm  4.25$ & $1451.25 \pm 163.74$ & $  2.97^{+  3.75}_{-  2.38}$ & XMMXCS J233607.6-535232.4 \\ 
0.300 & $ 91.43 \pm  4.92$ & $1745.27 \pm 143.40$ & $  3.95^{+  4.17}_{-  3.75}$ & XMMXCS J052215.8-481817.2 \\ 
0.222 & $ 91.20 \pm  4.36$ & $865.47 \pm 111.05$ & $  5.07^{+  5.19}_{-  4.96}$ & XMMXCS J022553.4-415448.4 \\ 
0.207 & $ 86.58 \pm  4.46$ & $825.73 \pm 113.98$ & $  6.08^{+  6.23}_{-  5.93}$ & XMMXCS J051016.7-451917.2 \\ 
0.383 & $ 86.38 \pm  4.27$ & $1775.89 \pm 260.44$ & $  6.21^{+  6.56}_{-  5.89}$ & XMMXCS J011443.1-412351.5 \\ 
0.654 & $ 85.16 \pm  5.11$ & $930.11 \pm 141.68$ & $  5.29^{+  5.60}_{-  5.00}$ & XMMXCS J023301.8-581928.5 \\ 
0.409 & $ 83.43 \pm  4.16$ & $767.95 \pm 135.16$ & $  6.94^{+  7.45}_{-  6.47}$ & XMMXCS J024038.2-594605.3 \\ 
0.114 & $ 82.41 \pm  3.37$ & $1147.89 \pm 195.46$ & $  5.24^{+  5.30}_{-  5.18}$ & XMMXCS J224622.0-524422.6 \\ 
0.540 & $ 80.43 \pm  7.59$ & $410.21 \pm  49.21$ & $  3.96^{+  4.81}_{-  3.37}$ & XMMXCS J234155.7-530843.5 \\ 
0.124 & $ 78.69 \pm  2.83$ & $1120.11 \pm 112.26$ & $  6.28^{+  6.36}_{-  6.20}$ & XMMXCS J014459.1-530113.7 \\ 
0.101 & $ 78.00 \pm  4.23$ & $1591.22 \pm 133.73$ & $  3.88^{+  3.94}_{-  3.81}$ & XMMXCS J214621.8-571719.3 \\ 
0.463 & $ 75.01 \pm  4.03$ & $1135.02 \pm 142.06$ & $  4.33^{+  4.51}_{-  4.17}$ & XMMXCS J040352.4-571939.7 \\ 
0.189 & $ 72.82 \pm  3.19$ & $567.63 \pm  70.24$ & $  6.67^{+  6.96}_{-  6.39}$ & XMMXCS J052548.9-471507.3 \\ 
0.246 & $ 71.23 \pm  4.94$ & $913.29 \pm 165.84$ & $  5.29^{+  5.34}_{-  5.24}$ & XMMXCS J212939.7+000516.9 \\ 
0.407 & $ 70.42 \pm  3.09$ & $982.90 \pm 136.59$ & $  4.84^{+  5.21}_{-  4.52}$ & XMMXCS J035415.2-590519.1 \\ 
0.168 & $ 66.85 \pm  4.18$ & $581.55 \pm  90.59$ & $  2.81^{+  3.52}_{-  2.29}$ & XMMXCS J232612.8-531858.4 \\ 
0.102 & $ 66.39 \pm  4.43$ & $529.56 \pm  71.17$ & $  4.57^{+  4.62}_{-  4.53}$ & XMMXCS J222353.0-013714.4 \\ 
0.274 & $ 61.95 \pm  2.66$ & $399.35 \pm  46.23$ & $  2.87^{+  3.30}_{-  2.53}$ & XMMXCS J231912.9-540457.7 \\ 
0.396 & $ 59.90 \pm  3.48$ & $495.71 \pm  73.44$ & $  4.18^{+  4.46}_{-  3.92}$ & XMMXCS J203049.5-563758.6 \\ 
0.547 & $ 59.51 \pm  3.67$ & $609.89 \pm 121.03$ & $  6.73^{+  7.59}_{-  5.99}$ & XMMXCS J034301.6-551835.5 \\ 
0.153 & $ 59.12 \pm  3.41$ & $527.02 \pm  64.94$ & $  3.75^{+  3.85}_{-  3.65}$ & XMMXCS J044956.6-444017.3 \\ 
0.140 & $ 57.74 \pm  2.84$ & $1022.11 \pm 175.99$ & $  3.87^{+  4.00}_{-  3.75}$ & XMMXCS J205556.3-545548.2 \\ 
0.614 & $ 56.95 \pm  3.80$ & $693.10 \pm 125.43$ & $  2.37^{+  3.00}_{-  1.95}$ & XMMXCS J231623.3-590432.4 \\ 
0.462 & $ 55.59 \pm  4.32$ & $856.54 \pm 124.87$ & $  5.32^{+  5.65}_{-  5.01}$ & XMMXCS J031715.7-593525.4 \\ 
0.599 & $ 52.66 \pm  3.22$ & $517.48 \pm  88.22$ & $  5.17^{+  5.86}_{-  4.61}$ & XMMXCS J023106.3-540349.9 \\ 
0.246 & $ 51.44 \pm  2.82$ & $885.71 \pm 126.48$ & $  2.66^{+  2.94}_{-  2.43}$ & XMMXCS J213004.1-002105.9 \\ 
0.386 & $ 49.95 \pm  2.83$ & $674.86 \pm 109.34$ & $  4.16^{+  4.82}_{-  3.63}$ & XMMXCS J231720.4-535734.5 \\ 
0.460 & $ 48.90 \pm  3.47$ & $436.30 \pm  53.15$ & $  4.56^{+  6.20}_{-  3.53}$ & XMMXCS J224549.8-525436.4 \\ 
0.131 & $ 48.53 \pm  3.20$ & $385.03 \pm  54.84$ & $  4.45^{+  4.56}_{-  4.34}$ & XMMXCS J203157.5-562430.2 \\ 
0.564 & $ 47.36 \pm  3.49$ & $577.55 \pm  72.13$ & $  2.81^{+  3.04}_{-  2.61}$ & XMMXCS J042226.4-514025.8 \\ 
0.393 & $ 46.96 \pm  3.27$ & $666.66 \pm 129.33$ & $  3.15^{+  3.30}_{-  3.00}$ & XMMXCS J003428.0-431854.2 \\ 
0.106 & $ 42.60 \pm  2.71$ & $895.62 \pm  97.66$ & $  3.47^{+  3.57}_{-  3.36}$ & XMMXCS J014030.7-543120.6 \\ 
0.106 & $ 42.11 \pm  3.61$ & $644.67 \pm  73.43$ & $  3.06^{+  3.15}_{-  2.98}$ & XMMXCS J003016.0-532513.6 \\ 
0.428 & $ 41.86 \pm  3.04$ & $615.71 \pm  85.38$ & $  5.02^{+  5.82}_{-  4.35}$ & XMMXCS J025720.9-573248.9 \\ 
0.389 & $ 39.18 \pm  2.70$ & $573.12 \pm  87.95$ & $  1.92^{+  2.55}_{-  1.55}$ & XMMXCS J232543.0-531635.8 \\ 
0.422 & $ 39.12 \pm  2.87$ & $554.71 \pm  95.61$ & $  2.56^{+  2.78}_{-  2.36}$ & XMMXCS J010030.2-474919.6 \\ 
0.584 & $ 38.90 \pm  3.91$ & $316.50 \pm  69.14$ & $  1.78^{+  2.23}_{-  1.49}$ & XMMXCS J034421.9-534042.5 \\ 
\end{tabular}
\end{table*}

\begin{table*}\caption*{continued.}\label{tab:chandra-tab}
    \centering
\begin{tabular}{|l|l|l|l|l|l|}
0.564 & $ 38.75 \pm  3.13$ & $723.56 \pm 108.57$ & $  2.12^{+  2.85}_{-  1.69}$ & XMMXCS J233330.2-521511.5 \\ 
0.218 & $ 38.46 \pm  2.44$ & $348.25 \pm  76.23$ & $  2.32^{+  2.43}_{-  2.21}$ & XMMXCS J003346.3-431729.7 \\ 
0.562 & $ 37.59 \pm  3.18$ & $216.50 \pm  59.47$ & $  3.32^{+  3.88}_{-  2.88}$ & XMMXCS J023209.8-574558.9 \\ 
0.373 & $ 37.20 \pm  2.91$ & $399.58 \pm  63.64$ & $  2.29^{+  2.90}_{-  1.89}$ & XMMXCS J041644.8-552506.6 \\ 
0.410 & $ 36.69 \pm  2.72$ & $374.30 \pm  93.59$ & $  4.40^{+  5.26}_{-  3.73}$ & XMMXCS J022728.2-405101.7 \\ 
0.416 & $ 35.96 \pm  2.87$ & $459.85 \pm  86.98$ & $  2.67^{+  3.35}_{-  2.21}$ & XMMXCS J011949.7-440434.5 \\ 
0.136 & $ 35.11 \pm  2.50$ & $341.51 \pm  49.72$ & $  2.51^{+  2.86}_{-  2.23}$ & XMMXCS J213027.0-000029.7 \\ 
0.310 & $ 34.66 \pm  2.84$ & $2850.61 \pm 245.68$ & $  3.85^{+  4.15}_{-  3.58}$ & XMMXCS J222314.6-013936.8 \\ 
0.422 & $ 33.60 \pm  2.92$ & $360.23 \pm  96.59$ & $  4.36^{+  5.25}_{-  3.69}$ & XMMXCS J054152.3-405236.4 \\ 
0.318 & $ 32.10 \pm  2.58$ & $430.69 \pm 107.72$ & $  2.31^{+  3.00}_{-  1.89}$ & XMMXCS J233644.6-534806.9 \\ 
\hline
\end{tabular}
\end{table*}

\end{document}